\documentclass[aps,prb,twocolumn,showpacs,preprintnumbers,amsmath,amssymb,superscriptaddress,floatfix,longbibliography]{revtex4-2}
\usepackage{graphicx}
\usepackage{hyperref}
\usepackage{amsfonts}
\usepackage{CJK}
\usepackage{float}
\newcommand\kv{\mathbf{k}}

\newcommand\Iv{\mathbf{I}}

\newcommand{\up}{{\uparrow}}
\newcommand{\down}{\downarrow}
\newcommand\ket[1]{|#1\rangle}
\newcommand\bra[1]{\langle#1|}

\begin{document}




\title{Topological phases in twisted Rashba superconductors}

\author{Conghao Lin}
\affiliation{Department of Physics, Xiamen University, Xiamen 361005, China}
 
\author{Xiancong Lu}\email{xlu@xmu.edu.cn}
\affiliation{Department of Physics, Xiamen University, Xiamen 361005, China}

\date{\today}

\begin{abstract}
  We study the topological properties of a twisted superconducting
  bilayer with spin-singlet pairings and Rashba spin-orbital coupling.
  By introducing the chirality basis, we obtain the effective
  odd-parity superconductors with the help of spin-orbital coupling.
  For the twisted bilayer with $d$-wave pairings, two non-Abelian
  topological phases with Chern number $C=-1$ and $C=-5$ are
  identified, and the analytical expressions for the boundary of
  non-Abelian phase are derived as well within the circular Fermi
  surface approximation. We perform numerical calculations at the
  twisted angle of Moir\'e lattice, which further verify the
  topological phase diagram from the effective odd-parity Hamiltonian.
  For the bilayer with $d$-wave and $s_{\pm}$-wave pairings, we reveal
  the second-order topological superconductor with Majorana zero mode on
  each corner, by analyzing the relative configuration of the pairing
  nodes of superconductors and the Fermi surface of normal state.  It
  is found that the regions of second-order topological phase are
  narrowed when the bilayer is twisted.
\end{abstract}


\maketitle


\section{INTRODUCTION}

The search for topological superconductors (TSCs) hosting Majorana
zero modes (MZMs) with non-Abelian exchange statistics is the focus of
attention in the community of condensed matter physics, due to the
potential application in topological quantum computation
\cite{qi.zh.11, sa.an.17}. Originally, the MZMs are predicted to exist
near the edges or within vortices of chiral $p$-wave superconductor
(SC) \cite{re.gr.00,kita.01,ivan.01}. However, $p$-wave superconductor is rare in
nature. A more practical route toward TSCs is to engineer a
heterostructure using the conventional even-parity SCs and
topologically nontrivial materials. For example, deposit the $s$-wave SC on the
surface of a topological insulator (TI) \cite{fu.ka.08} or bring it in
proximity to a semiconductor with Rashba spin-orbit coupling (SOC)
\cite{sa.lu.10,or.re.10,alic.10,alic.12}.

Recently, the monolayer cuprate Bi$_2$Sr$_2$CaCu$_2$O$_{8+\delta}$
(Bi2212) has been successfully realized in experiment, which has a
high transition temperature close to bulk
samples~\cite{yu.ma.19,zh.po.19}. Meanwhile, plenty of novel phenomena
has been found in the twisted van der Waals materials
\cite{bi.ma.11,ca.fa.18a,ca.fa.18b,wu.lo.19,an.ma.20}. Inspired by
these achievements, Can \textit{et
  al.} propose to realize the TSCs in twisted bilayer
cuprate~\cite{ca.tu.21}. By stacking two monolayers of cuprate SC
together and twisting them at a large angle (close to $45^\circ$), a
time-reversal symmetry breaking $d+id$ superconducting phase is argued
to emerge,
which is fully gapped and topologically nontrivial
\cite{go.ku.04,ya.qi.18}. This promising proposal stimulates some
experimental and theoretical works to investigate the pairing symmetry
and topology of twisted bilayer cuprate
\cite{zh.cu.23,le.le.21,zh.wa.23,wa.zh.23,so.zh.22,lu.se.22,li.zh.23}. However,
a consensus about the nature of SC in this system has not been reached
yet.


On the other hand, the Chern numbers in the proposal by Can \textit{et
  al.} are always even \cite{ca.tu.21}, due to the nature of singlet
SC pairings. In this case, the chiral Majorana edge modes come into
pairs and therefore cannot form the non-Abelian Fermions.
A way to overcome this
shortcoming is to incorporate the SOC into the platform of cuprate to
lift the spin degeneracy \cite{alic.12,sa.an.17}. There are already
theoretical proposals: deposit the twisted bilayer
cuprate on the surface of a strong TI \cite{me.sa.22}, or grow it in proximity
to the semiconductor $Bi_2O_2Se$, which has a large Rashba SOC and a matching
lattice constant \cite{ma.ya.22}. Furthermore, a hidden Rashba SOC has
been demonstrated in the well-studied cuprate Bi2212
\cite{go.li.18}. The origin of this hidden SOC is attributed to the locally
noncentrosymmetric crystal structure, in which the inversion symmetry
is broken locally but not globally
\cite{fi.lo.11,ma.si.12,zh.li.14,lu.se.21,fi.si.22}.

However, a systematic analysis of non-Abelian topological order in
twisted bilayer superconductors with SOC has not been performed
yet. This paper aims to fill this gap. Previous work by Sato
\textit{et al.} has explored non-Abelian topological orders in a
system of single-layer $d+id$ SC with Rashba SOC and a Zeeman magnetic
field \cite{sa.ta.10,sa.fu.10}. We will generalize that work to a
twisted bilayer superconducting system, within the approximation of circular Fermi
surface (FS) \cite{ca.tu.21,vo.zh.21}. When both layers have $d$-wave
SC pairings, we will identify the boundary of non-Abelian topological
phase. Furthermore, we will investigate the high-order topology in the
system where two layers have different SC pairings, with one
layer being $d$-wave and the other layer $s_{\pm}$-wave.

This paper is organized
as follows.  In Sec.~\ref{sec:model}, we introduce the model
Hamiltonian for the twisted bilayer, in both momentum and real
spaces. In Sec.~\ref{sec:transformation}, a similarity transformation
is introduced, which transforms the even-parity SC with Rashba SOC
into an effective odd-parity SC. In Sec.~\ref{sec:d+id}, the
non-Abelian topological phases are discussed in the twisted $d$-wave
bilayer. In Sec.~\ref{sec:d+is}, we investigate the high-order
topological SC in the twisted bilayer with $d$- and $s_{\pm}$-wave
pairings. Finally, brief conclusions are presented in
Sec.~\ref{sec:conclusion}.

\begin{figure}
\includegraphics[width=1.0\columnwidth]{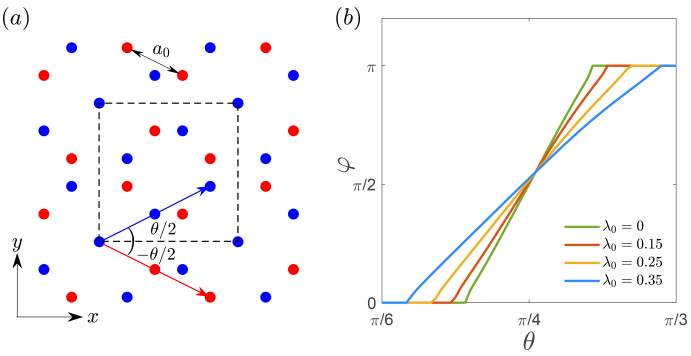}
\caption{(Color online) The panel $(a)$: Twisted bilayer at a
  commensurate angel $\theta=53.13^\circ$. The panel $(b)$: The
  relative phase $\varphi$ between two superconducting layers as a
  function of twisted angle $\theta$, for various magnitudes of Rashba
  SOC $\lambda_0$. The presence of Rashba SOC enlarges the
  time-reversal symmetry breaking region. The values of other
  parameters are chosen as $t=1$,
  $\mu=-3.5$, $\Delta_{d0}=0.5$, and $t^0_z=0.25$.}
\label{fig:rela_phas}
\end{figure}

\section{Model Hamiltonian}\label{sec:model}

We consider a bilayer of spin-singlet superconductor, in which one
superconducting layer is twisted by an angle with respect to the other
one. The spin-orbit coupling in each layer is induced by proximity to
two-dimensional materials or semiconductors \cite{ma.ya.22}, or by considering a
locally noncentrosymmetric structure \cite{fi.si.22}.
The effect of magnetic field perpendicular to the bilayer is
also taken into account. In the framework of BCS
mean-field theory, the Hamiltonian of bilayer can be written as 
\begin{eqnarray}\label{Ham}
H &=& \sum_{\kv_l,\sigma}\xi(\mathbf{k}_l)
        c_{\mathbf{k}_l\sigma}^{\dagger}c_{\mathbf{k}_l\sigma}
      + \sum_{\mathbf{k}_l,\sigma,\sigma'}
        \mathbf{L}(\mathbf{k}_l)\cdot\boldsymbol{\sigma}_{\sigma\sigma'}
        c_{\mathbf{k}_l\sigma}^{\dagger} c_{\mathbf{k}_l\sigma'}\nonumber\\
   && + \sum_{\mathbf{k}_a, \kv_b, \sigma}t_z(\mathbf{k}_a,\mathbf{k}_b)
        ( c_{\mathbf{k}_a\sigma}^{\dagger}c_{\mathbf{k}_b\sigma} + h.c.)\nonumber\\
   && + \sum_{\mathbf{k}_l,l} \Big[ \Delta_l(\mathbf{k}_l)
       c_{\mathbf{k}_l\uparrow}^{\dagger}c_{-\mathbf{k}_l\downarrow}^{\dagger}
      + \Delta_l^*(\mathbf{k}_l)
       c_{\mathbf{-k}_l\down}c_{\mathbf{k}_l\up} \Big] \nonumber\\
   && - h_z\sum_{\mathbf{k}_l,\sigma,\sigma'}(\sigma_{z})_{\sigma\sigma'}
       c_{\mathbf{k}_l,\sigma}^{\dagger}c_{\mathbf{k}_l,\sigma'}.
\end{eqnarray}
Here, the operator $c^{\dagger}_{\mathbf{k}_l,\sigma}$
($c_{\mathbf{k}_l,\sigma}$) creates (annihilates) an electron in layer
$l$ ($l=a,b$ is a layer index) with momentum $\mathbf{k}_l$ and spin
$\sigma$.
We follow the convention that the $a$ and $b$ layers are rotated by
angles $\frac{\theta}{2}$ and $-\frac{\theta}{2}$, respectively, with
respect to a reference plane that is unrotated; see Fig. \ref{fig:rela_phas}(a) for an
example. The momentum $\kv$, associated with the unrotated plane,
is connected to $\kv_a$ and $\kv_b$ by
\begin{equation}
 \kv_a= R(\frac{\theta}{2})\kv, \quad \kv_b= R(-\frac{\theta}{2})\kv
\end{equation}
in which the $R$ is two dimensional rotation matrix,
\begin{equation}
R(\frac{\theta}{2})=
    \begin{pmatrix}
    \cos\frac{\theta}{2} & -\sin\frac{\theta}{2}\\
    \sin\frac{\theta}{2} &  \cos\frac{\theta}{2}
    \end{pmatrix}
\end{equation}
The single-particle dispersion in each layer is
$\xi(\mathbf{k}_{l})=-2t(\cos k_{l,x}+\cos k_{l,y})-\mu$, in which $t$
is the hopping amplitude and $\mu$ the chemical potential.
In the remainder of this paper, we set $t$ as the
energy unit of the system, \textit{i.e.}, $t=1$.
The Rashba SOC is described by
$\mathbf{L}(\kv_l)\cdot\boldsymbol{\sigma}$, in which
$\mathbf{L}(\kv_l)=2\lambda_0 (\sin k_{l,y}, -\sin k_{l,x}, 0)$ is a
vector, $\lambda_0$ is the magnitude of SOC, and $\boldsymbol{\sigma}$
is Pauli matrixes with components $\sigma_i(i=0,x,y,z)$. Two layers
are coupled by a spin-independent single-particle tunneling term
$t_z(\mathbf{k}_a,\mathbf{k}_b)$, which in general 
depends on both momentums $\kv_a$ and $\kv_b$ \cite{so.zh.22, vo.zh.21}. The
Zeeman magnetic field $h_z$ perpendicular to the layers is also
included in the Hamiltonian. We consider a spin-singlet SC in each
monolayer. The gap function for a $d$-wave SC is
$\Delta_d(\kv)=2\Delta_{d0}(\cos k_x-\cos k_y)$, and for the extended
$s$-wave SC is $\Delta_s(\kv)=\Delta_{s0}+2\Delta_{s1}(\cos k_x+\cos
k_y)$.
In this paper, we disregard the spin-triplet SC that might be induced
by the Rashba SOC, as the amplitude of spin-triplet SC is typically
small \cite{yo.ya.16}.
Due to the interlayer tunneling, there might exist a relative phase
difference $\varphi$ between two SC layers, \textit{i.e.},
$\Delta_a(\kv_a)=\Delta_d(\kv_a)$ and
$\Delta_b(\kv_b)=e^{i\varphi}\Delta_d(\kv_b)$, which gives rise to the
time-reversal symmetry breaking SC phase
\cite{ya.qi.18,ca.tu.21,vo.wi.23a,vo.wi.23b}.

By introducing the Nambu spinor
$\Psi_\kv^{\dagger}= ( c^{\dagger}_{\kv_a\up}, c^{\dagger}_{\kv_a\down},
         c_{-\kv_a\up},c_{-\kv_a\down},
         c^{\dagger}_{\kv_b\up}, c^{\dagger}_{\kv_b\down},
         c_{-\kv_b\up},c_{-\kv_b\down} )$,
The Hamiltonian (\ref{Ham}) can be written in the Bogoliubov-de-Gennes
(BdG) formalism,
\begin{eqnarray}
H = \frac{1}{2} \sum_\kv \Psi^{\dagger}_{\kv} H_{BdG}(\kv)\Psi_{\kv},
\end{eqnarray}
in which $H_{BdG}(\kv)$ is given by
\begin{eqnarray}\label{HBdG}
H_{BdG}(\kv)=
\begin{pmatrix}
H_a(\kv_a)         &      T(\kv_a,\kv_b)\\
 T(\kv_a,\kv_b)    &      H_b(\kv_b)
\end{pmatrix}.
\end{eqnarray}
The $H_{l}(\kv_l)$ is the Hamiltonian for the single $l$-layer,
\begin{eqnarray}
H_{l}(\kv_l)=
\begin{pmatrix}
H_0(\kv_l)                   &  i\Delta_l(\kv_l)\sigma_y \\
-i\Delta_l^*(\kv_l)\sigma_y  &          -H_0^T(-\kv_l)
\end{pmatrix}
\end{eqnarray}
with
\begin{eqnarray}\label{H0}
H_0(\kv_l)= \xi(\kv_l)\sigma_0 - h_z\sigma_z +
            \mathbf{L}(\kv_l)\cdot\boldsymbol{\sigma}
\end{eqnarray}
being the normal-state
Hamiltonian of $l$-layer. The $T(\kv_a,\kv_b)$ term in
Eq. (\ref{HBdG}) describes the interlayer tunneling, which is a matrix
given by
\begin{eqnarray}
  T(\kv_a,\kv_b)=t_z(\kv_a,\kv_b)\sigma_0\tau_z,
\end{eqnarray}
in which $\boldsymbol{\tau}$ are the Pauli matrices in Nambu
particle-hole notation.

The interlayer tunneling $t_z(\kv_a,\kv_b)$ couples the momentums in
two layers, whose structure is complex in general
\cite{mo.ko.13,kosh.15,bi.ma.11}. For square lattices with interlayer
hopping decaying exponentially in real space, a commonly used
approximation for $t_z(\kv_a,\kv_b)$ is to treat it as a constant
\cite{ca.tu.21,vo.wi.23a,vo.wi.23b,vo.zh.21,ya.qi.18,tu.pl.22,tu.la.22},
\textit{i.e.}, keep only the momentum-conserving parts. We denote this constant as
$t_z(\kv_a,\kv_b)=t_z^0$ when adopting this
approximation in this paper.
For layered superconductors like cuprates, $d$-wave superconductivity
occurs in the $CuO$ planes, and the interlayer hopping is
significantly smaller than the intralayer hopping. Numerical studies
based on the bilayer Hubbard model have verified that a small
interlayer tunneling has minimal effect on the in-plane
superconducting correlation and gap functions in a large doping
region \cite{iw.ya.22,sc.ca.94}.
To simplify our analysis, we do not consider the stability of the SC phase against
interlayer tunneling, but instead treat the interlayer coupling as a
minor perturbation that does not alter the magnitude of
the SC order parameter in each layer \cite{ca.tu.21,ha.tu.22}.
Hence the relative phase
$\varphi$ between the SC gap functions of two layers only depends on
the twisted angle $\theta$, which can be obtained by minimizing
the ground state energy ($T=0$) \cite{ya.qi.18,ca.tu.21},
\begin{equation}
 E_{GS}(\theta)=\sum_{n}\sum_{\kv}E_n(\kv,\theta),
\end{equation}
where $E_n$ is the energy eignvalues by diagonalizing $H_{BdG}$ in Eq.
(\ref{HBdG}), and index $n$ runs four occupied energy bands.

\subsection{Hamiltonian in real space}
The tight-binding lattice Hamiltonian for bilayer SC in real space
representation reads
\begin{eqnarray}
H = H^{(a)} + H^{(b)} + H_{\bot},  
\end{eqnarray}
where the single-layer Hamiltonian $H^{(l)} (l=a,b)$ is 
\begin{eqnarray}
H^{(l)} & =& -t\sum_{\langle ij\rangle \sigma}c^{\dagger}_{il\sigma}c_{jl\sigma}
         - h_z\sum_i(c^{\dagger}_{il\uparrow}c_{il\uparrow} -
                    c^{\dagger}_{il\downarrow}c_{il\downarrow})
         \nonumber\\
      && +\sum_{\langle ij\rangle }\Big(
          \Delta_{ij,l}c^{\dagger}_{il\uparrow}c^{\dagger}_{jl\downarrow}
         + h.c. \Big)
         -\mu\sum_{i\sigma}n_{il\sigma}\nonumber\\
      && - \lambda_0\sum_i \Big[ (   c_{i-\hat{x},l\downarrow}^\dagger c_{il\uparrow}
                                - c_{i+\hat{x},l\downarrow}^\dagger c_{il\uparrow} ) \nonumber\\
&&                          + i(   c_{i-\hat{y},l\downarrow}^\dagger c_{il\uparrow}
                                - c_{i+\hat{y},l\downarrow}^\dagger c_{il\uparrow} )
   + h.c. \Big],
\end{eqnarray}
in which $i,j$ are the site indexes for a square lattice (each layer
has its own), and $\hat{x}(\hat{y})$ are the lattice unit vector along
the $x(y)$ direction in each layer. The $\Delta_{ij,l}$ is the complex
SC order parameter on the bond connecting sites $i$ and $j$. The
Hamiltonian of interlayer tunneling $H_\bot$ is given by
\begin{eqnarray}
 H_{\bot} =  -\sum_{ij,\sigma}g_{ij}c^{\dagger}_{ia\sigma}c_{jb\sigma},
\end{eqnarray}
where $g_{ij}$ is the amplitude of interlayer tunneling. In this
paper, we take a simplified form for $g_{ij}$ as in
Ref. \cite{ca.tu.21}, which decays exponentially with distance:
\begin{equation}\label{gij}
g_{ij}=g_{0}e^{-(r_{ij}-c)/\rho}
\end{equation}
where $r_{ij}$ is the distance between the site $i$ in $a$-layer and
site $j$ in $b$-layer, $r_{ij}=\sqrt{c^2+d_{ij}^2}$, with $d_{ij}$
being the in-plane separation between sites $i$ and $j$ and $a$ being
the interlayer distance. The parameter $\rho$ in Eq. (\ref{gij})
denotes a phenomenological decay constant \cite{ca.tu.21}. In
calculations, we use $c=2.2,\rho=0.4$, which are in units of the lattice
constant $a_0$ of square lattice in each layer, corresponding to the case of weak
interlayer tunneling \cite{ca.tu.21}.


When the twist angle satisfies the condition $\theta=2\arctan(n/m)$,
with $n$, $m$ being integers, the bilayer are commensurate and thus
forms a periodic Moir\'e lattice, whose unit cell contains
$2(m^2+n^2)$ lattice sites. Based on the Moir\'e lattice, the Bloch
representation of wave functions can be used, such that the Chern
number and edge states can be calculated numerically \cite{ca.tu.21,tu.la.22}.

\section{The transformation to effective odd-parity superconductor}
\label{sec:transformation}

The topological properties of model (\ref{Ham}) are determined by the
structure of superconducting pairings and the FS of the
normal state.
Note that the Bogoliubov Fermi surface (BFS) does not exist in this model when
both magnetic field and Rashba SOC are present, and therefore there
is no need to consider the effect of BFS on the system's topology.
For simplicity, we assume a circular Fermi surface for
the normal state of each layer \cite{ca.tu.21,vo.zh.21}, so that the
normal-state Hamiltonian in Eq. (\ref{H0}) does not depend on the
twisted angle and can be written as $H_{0}(\kv)$. This approximation
is valid when FS approaching the center of Brillouin zone (BZ), for the energy
spectrum of $H_{0}(\kv)$ is invariant under rotation in this case.

We then introduce a unitary transformation $S$ to express the BdG Hamiltonian
(\ref{HBdG}) on the bonding and anti-bonding basis, 
\begin{equation}\label{s}
S=\frac{1}{\sqrt{2}}
\begin{pmatrix}
\Iv_{4\times4}&-\Iv_{4\times4}\\
\Iv_{4\times4}&\Iv_{4\times4}
\end{pmatrix},
\end{equation}
in which $\Iv_{4\times4}$ denotes a $4\times4$ identity matrix. After
transformation, the pairing terms of the upper and lower layers are
mixing and the Hamiltonian (\ref{HBdG}) takes a new
form,
\begin{equation}\label{HpBdG}
H_{BdG}^{'}=SH_{BdG}S^\dagger=
\begin{pmatrix}
H_{+}&B\\
B&H_{-}
\end{pmatrix}
\end{equation}
Here, the matrix $H_{\pm}$ and $B$ are given by
\begin{equation}\label{Hpm}
H_{\pm}=
\begin{pmatrix}
H_0^{\pm}(\kv)            &  i\Delta_{a+b}\sigma_y\\
-i\Delta^*_{a+b}\sigma_y  &  [-H_0^{\pm}(-\kv) ]^T
\end{pmatrix}
\end{equation}

\begin{equation}\label{B}
B=
\begin{pmatrix}
       0                &    i\Delta_{a-b}\sigma_y\\
-i\Delta_{a-b}^*\sigma_y   &      0
\end{pmatrix},
\end{equation}
in which $H_0^{+}(H_0^{-})$ is the normal state Hamiltonian for the
bonding (anti-bonding) band,
\begin{equation}\label{H0pm}
H_0^{\pm}(\kv)=
\begin{pmatrix}
\epsilon_{\pm}(\mathbf{k})-h_z & \lambda(\mathbf{k}) \\
\lambda^{*}(\mathbf{k})  &\epsilon_{\pm}(\mathbf{k})+h_z 
\end{pmatrix}
\end{equation}
with
\begin{eqnarray}
  \epsilon_{\pm}(\mathbf{k})&=&\xi(\kv) \mp t_z^0\\
  \lambda(\mathbf{k})&=&2\lambda_{0}(\sin k_y + i\sin k_x )
\end{eqnarray}
and $\Delta_{a+b} (\Delta_{a-b})$ is the pairing mixing term of two layers
\begin{eqnarray}\label{apmb}
  \Delta_{a\pm b}=\frac{1}{2} \Big(
                     \Delta_a(\kv_a)\pm\Delta_b(\kv_{b}) \Big).
\end{eqnarray}

As shown by Sato \textit{et al.}, a single layer $d+id$ superconductor
can be mapped to spinless chiral $p$-wave superconductor
\cite{sa.ta.10, sa.ta.09}, with the help of Rashba SOC
interaction. Here, we generalize this idea to the twisted bilayer
system. The key step is to introduce the chirality basis, on which the normal state Hamiltonian $H_0^\pm(\kv)$
in Eq. (\ref{H0pm}) can be diagonalized. In particular, the following
chirality transformation is used to diagonalize the normal state
Hamiltonian for a single layer \cite{sa.ta.10},
\begin{equation}\label{U}
U(\mathbf{k})=\frac{1}{\sqrt{2\eta(\kv)(\eta(\kv)+h_z)}}
\begin{pmatrix}
\lambda(\kv)  & -\eta(\kv)-h_z\\
\eta(\kv)+h_z & \lambda^*(\kv)
\end{pmatrix}
\end{equation}
with
\begin{equation}\label{eq13}
\eta(\kv)=\sqrt{h_z^2+|\lambda(\kv)|^2}.
\end{equation}
For the twisted bilayer Hamiltonian (\ref{HpBdG}), we construct a
similar transformation $G$ to the chirality basis based on the matrix $U(\kv)$,
\begin{equation}
G= 
\begin{pmatrix}
g & 0\\
0 & g
\end{pmatrix},\quad
g = 
\begin{pmatrix}
U^{\dagger}(\mathbf{k}) & 0\\
0&U^{T}(-\mathbf{k})
\end{pmatrix}.
\end{equation}
The Hamiltonian (\ref{HpBdG}) in the chirality basis representation is
written as
\begin{equation}\label{wHBdG}
  \widetilde{H}_{BdG}=G H^{'}_{BdG}G^\dagger=
    \begin{pmatrix}
     \widetilde{H}_+ &   \widetilde{B}    \\
     \widetilde{B}   &   \widetilde{H}_-
  \end{pmatrix}.
\end{equation}
Here the matrix $\widetilde{H}_\pm$ and $\widetilde{B}$ are given by
\begin{eqnarray}
  \widetilde{H}_{\pm}=g H_{\pm} g^\dagger=
  \begin{pmatrix}
     \epsilon_{\pm}(\mathbf{k})+\eta(\kv)\sigma_z  &   \widetilde{\Delta}_{+}\\
     (\widetilde{\Delta}_{+})^\dagger    & -\epsilon_{\pm}(\mathbf{k})-\eta(\kv)\sigma_z  
  \end{pmatrix},  
\end{eqnarray}  
\begin{eqnarray}\label{wB}
   \widetilde{B}=g B g^\dagger=
  \begin{pmatrix}
     0       &   \widetilde{\Delta}_{-}\\
     (\widetilde{\Delta}_{-})^\dagger    & 0  
  \end{pmatrix}, 
\end{eqnarray}  
in which $\widetilde{\Delta}_+(\widetilde{\Delta}_{-})$ are the effective gap function
\begin{eqnarray}\label{wDelta}
  \widetilde{\Delta}_\pm= \frac{1}{\eta(\kv)}
  \begin{pmatrix}
     \Delta_{a\pm b} \lambda^*(\kv)  &  h_z \Delta_{a\pm b}       \\
     -h_z \Delta_{a\pm b}  &  \Delta_{a\pm b} \lambda(\kv)
  \end{pmatrix}.
\end{eqnarray} 
Therefore, the odd-parity pairings (\textit{e.g.},
$\Delta_{a\pm b} \lambda^*(\kv)$) are induced in the chirality basis
due to the Rashba SOC interaction.

The matrix $\widetilde{H}_+$ ($\widetilde{H}_-$) has the same
structure as the Hamiltonian of single layer Rashba superconductors
discussed in Ref. \cite{sa.ta.10}, except that the even-parity pairing
$\Delta_{a\pm b}$ (see Eq. (\ref{apmb})) depends on the twisting angle
$\theta$ between two layers. The matrix $\widetilde{B}$ in
Eq. (\ref{wB}) describes the pairings between bonding and anti-bonding
bands, which becomes zero if the twisted angle $\theta=0$ and
the pairings of two layer are exactly the same $\Delta_a=\Delta_b$.

\begin{figure}[tp]
\includegraphics[width=0.96\columnwidth]{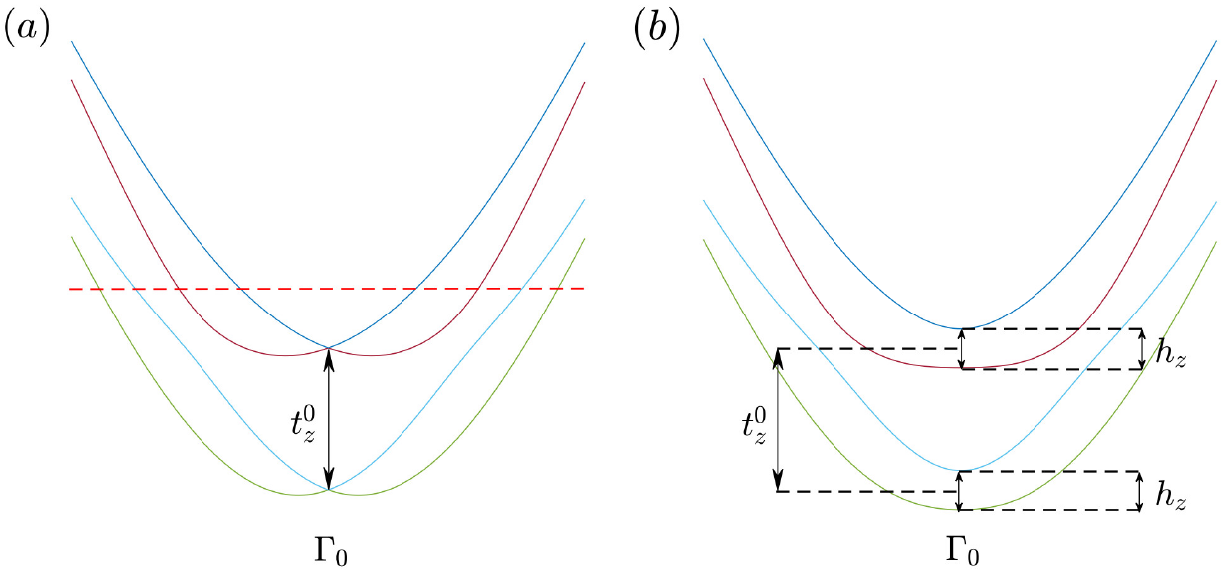}
\caption{(Color online) Schematic picture for the band structure of normal state. The
  energy bands are plotted close to the center of BZ, where the
  approximation of circular FS is valid. The Zeeman magnetic field
  $h_z$ is zero in panel $(a)$, while is nonzero in panel $(b)$.  At
  the point $\Gamma_0=(0,0)$, the degeneracy of energy bands is
  lifted up by the inter-layer hopping $t_z^0$ and the Zeeman field
  $h_z$.}
\label{band_NS}
\end{figure}

\section{Twisted bilayer with $d$-wave pairings}
\label{sec:d+id}

\subsection{The condition for gap close}

Monolayer $d$-wave SC has four Dirac nodes in the BZ, originating from
the intersection of SC nodal lines and the FS of normal state.  When
two monolayer SCs are stacked together and twisted at a certain angle,
the four Dirac nodes may be gapped out, inducing a nonzero Berry
curvature in the vicinity of FS \cite{ca.tu.21,vo.wi.23a,vo.wi.23b}.
Similarly, the topology of the Rashba bilayer SC is related to the nodal
lines of SC gap function and the structure of normal state FS.

In Fig. \ref{fig:rela_phas}(b), the relative phase $\varphi$ between
two superconducting layers are plotted as a function of twisted angle
$\theta$, for various values of Rashba SOC.
When the twisted angle $\theta=0^\circ$, there is no phase difference
between the $d$-wave SCs of two layers ($\varphi=0$). Therefore, the
bilayer system is still a nodal SC with gapless excitation, in which the
topological invariant is not well-defined in a strong sense
\cite{sa.fu.10}.
As $\theta$ increases from $0^\circ$, the relative phase $\varphi$ remains at zero. 
Once $\theta$ is larger than a critical value, a non-trivial phase
difference ($\varphi\neq 0, \pi$)
emerges, similar to previous results based on the BCS mean-field
theory \cite{ca.tu.21, vo.wi.23a,vo.wi.23b}. Interestingly, the presence of
Rashba SOC actually enlarges the region of $\theta$ with nontrivial
phase difference, as shown in Fig. \ref{fig:rela_phas}(b).
Note that a nontrivial phase difference cannot be induced by the twisted
angle for the bilayers with conventional $s$-wave pairings.
The effective gap function $\widetilde{\Delta}_\pm$
for the twisted $d$-wave bilayer depends on
$\Delta_{a\pm b}=\Delta_d(\kv_a)\pm e^{i\varphi}\Delta_d(\kv_b)$; see
Eqs.  (\ref{apmb}) and (\ref{wDelta}). When $\varphi$ is nontrivial,
the time-reversal symmetry is broken, so that $\Delta_{a\pm b}$ is gapped in BZ but still has
nodes at the time-reversal invariant (TRI) points (such as
$\Gamma_0=(0, 0)$).
In this case, if the FS of normal state crosses over the $\Gamma_0$
point, the energy band gap will close and reopen, implying the
possibility of topological phase transition.

A schematic picture for the band structure of normal state close to
the center $\Gamma_0$ of BZ is shown in Fig. \ref{band_NS}, where the
approximation of circular FS is valid. As shown in
Fig. \ref{band_NS}(a) where the Zeeman field is zero, the energy bands
are separated into bonding and anti-bonding bands due to the
interlayer hopping term, and the Rashba SOC term breaks the degeneracy of
spin for each band. However, the spin keeps degenerate at the
$\Gamma_0$ point, for the spatial asymmetry of Rashba SOC. As the Fermi
level varies, the spin-momentum locked FS emerges or disappears in
pairs; see the red dashed line in Fig. \ref{band_NS}(a) for
example. If applying an external Zeeman field $h_z$, the degeneracy of
spin at $\Gamma_0$ point can be further lifted, as shown in
Fig. \ref{band_NS}(b), such that there are four separated
bands in this case.
The condition for the normal state FS to cross the
$\Gamma_0$ point (the gap closing condition) can be obtained
straightforwardly,
\begin{equation}\label{eq:anal_boun}
\mu=-4 \pm t_z^0 \pm h_z
\end{equation}
in which $\xi(\Gamma_0)=-4-\mu$ is used for only the nearest
neighboring hopping in each layer is taken into account.

\subsection{The non-Abelian phase boundary}
\label{sec:d+id:B}

It is shown that a spinless odd-parity superconductor can be obtained
from $\widetilde{H}_\pm$ by integrating out fermion fields for the
high-energy massive band \cite{sa.ta.10}, which gives rise to the
non-Abelian topological order. When $|\lambda(\kv)| \gg h_z$, the
off-diagonal terms in $\widetilde{\Delta}_{\pm}$ are negligibly small
compared to the diagonal terms (odd-parity pairings), therefore, the
Hamiltonian $\widetilde{H}_{BdG}(\kv)$ in Eq. (\ref{wHBdG}) describes
an effective odd-parity SC in this case. The
$\widetilde{H}_{BdG}(\kv)$ with odd-parity SC has the following
symmetry,
\begin{equation}\label{Pi}
\Pi^{\dagger} \widetilde{H}_{BdG}(\kv) \Pi = \widetilde{H}_{BdG}(-\kv), \quad
\Pi = s_0 \otimes \tau_z \otimes \sigma_0,
\end{equation}
in which $s_0$ is the $2\times 2$ unit matrix in the space of
bonding and anti-bonding bands.
Combining the $\Pi$ symmetry with the particle-hole
symmetry, the topological criterion developed in Ref. \cite{sato.10}
can be generalized to the twisted bilayer system: the topology of odd
parity SC is determined by the signs of normal state dispersions at
the TRI points.

The Chern number can be calculated by integrating the field strength
$\mathcal{F}(\kv)$ of Berry connection $\mathcal{A}(\kv)$
over the whole BZ $T^2$, or equivalently by integrating $\mathcal{A}(\kv)$ along
the boundary of half BZ $\partial T^2/2$, 
\begin{equation}\label{eq:Chern}
C = \frac{1}{2\pi} \int_{T^2} d^2k \mathcal{F}(\kv)
  = \frac{1}{\pi}\oint_{\partial T^2/2}dk_i\mathcal{A}_i(\kv)
\end{equation}
with the Berry connection being defined as
\begin{equation}
\mathcal{A}_j (\kv) = i \sum_{E_n<0} \bra{u_n(\kv)} \partial_{k_j} \ket{u_n(\kv)}.
\end{equation}
For the odd-parity superconductor, the line integral in
Eq. ({\ref{eq:Chern}}) can be separated into two parts \cite{sato.10}
\begin{eqnarray}\label{eq16}
  C=w[L_{12}]-w[L_{34}],\\
  w[L_{ij}]=\frac{1}{\pi}\oint_{L_{ij}}dk_i\mathcal{A}_i(\kv),
\end{eqnarray}
in which $L_{ij}$ denotes a closed path passing through the TRI
momenta $\Gamma_i$ and $\Gamma_j$. It is interesting that the parity
of Chern number is related to the structure of Fermi surface by \cite{sato.10}
\begin{equation}\label{eq:C}
(-1)^{C}=\prod_{m}\prod_{i=0,1,2,3}sgn[\widetilde{\varepsilon}_m(\Gamma_{i})],
\end{equation}
in which $\Gamma_i(i=0,1,2,3)$ is the four TRI momenta for the square lattice,
$\widetilde{\varepsilon}_m(\kv)$ is the dispersions of normal
state for Hamiltonian $\widetilde{H}_{BdG}$ in Eq. (\ref{wHBdG}),
\begin{eqnarray}
\widetilde{\varepsilon}_m(\kv)=\xi(\kv) \mp t_z^0 \pm \eta(\kv)
\end{eqnarray}
in which the $m$ has four choices (four bands).

In this paper, we consider the parameter region in which the
approximation of circular FS for normal state is valid
\cite{ca.tu.21,vo.zh.21}, \textit{i.e.}, the FS is not far away from
the center of BZ. Therefore, only the first TRI point
$\Gamma_0=(0,0)$ is needed to take into account in the criteria of
Eq. (\ref{eq:C}). If the FSs enclose the $\Gamma_0$ point odd times,
the Chern number will be an odd number, corresponding to a non-Abelian
topological phase.
By varying the value of $\mu$, the number of FSs enclosing the
$\Gamma_0$ point will change; see the schematic band structure in
Fig. \ref{band_NS}. The boundary for the non-Abelian topological phase
can be obtained by substituting from
$\widetilde{\varepsilon}_m(\Gamma_0)$ into Eq. (\ref{eq:C}), which
turns out to be exactly the same as the gap-closing condition in
Eq. (\ref{eq:anal_boun}).

\begin{figure}
\includegraphics[width=0.96\columnwidth]{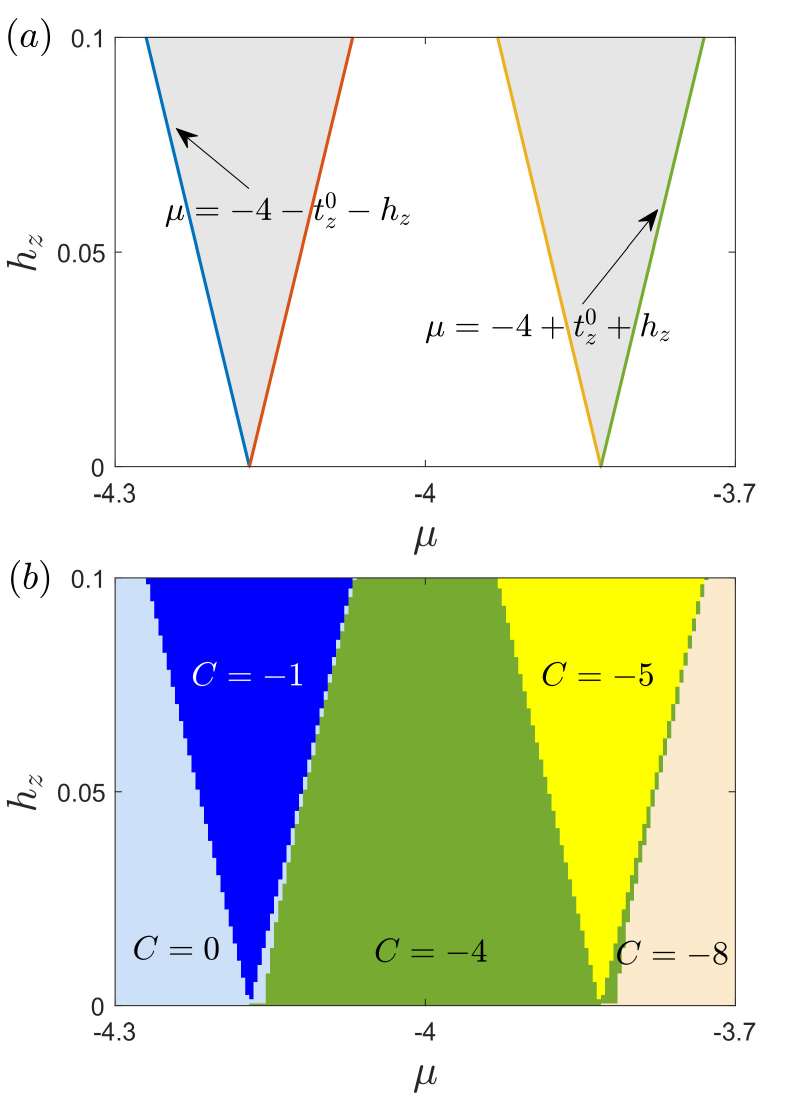}
\caption{Non-Abelian topological phase diagram for the twisted
  Rashba SC with $d$-wave pairings. Panel $(a)$ is the analytical
  phase boundaries obtained in Sec. \ref{sec:d+id:B}. Panel (b) is
  the numerical results of Chern number, which are calculated using
  the Moir\'e lattice at a twisted angle $\theta=43.6^\circ$. The
  values of parameters in the calculations are
  $t=1,\lambda_0=0.25,g_0=0.25,\Delta_{d0}=0.5$.}
\label{fig:chern}
\end{figure}

In Fig. \ref{fig:chern}(a), we plot these four lines of phase
boundary. The gray regions between two lines may be the non-Abelian
topological phase, for there are odd number of FSs enclosing the
$\Gamma_0$ point. Comparing with the phase diagram of single layer
$d+id$-wave SC in Ref. \cite{sa.ta.10}, one can see that the twisted
bilayer system has a new non-Abelian phase with a different Chern
number. This results from the doubled FSs of normal state by the
interlayer hopping term. We will compute the value of Chern number in the
following section.

\begin{figure}
\includegraphics[width=0.92\columnwidth]{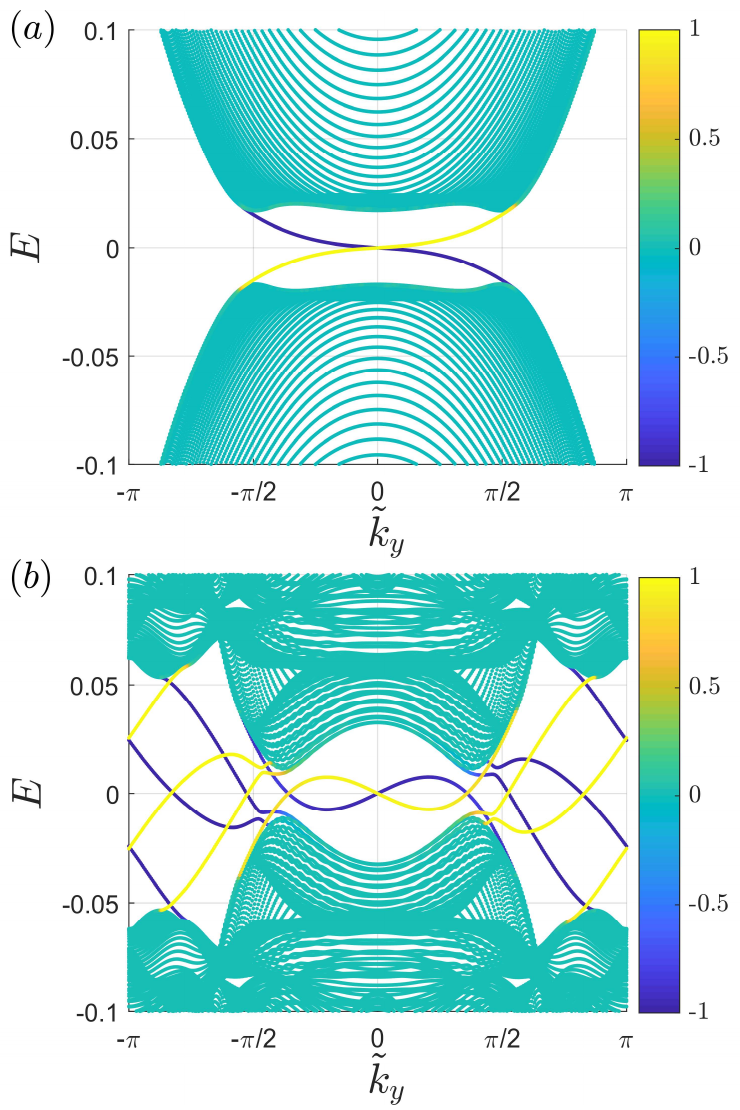}
\caption{Spectrum of Moir\'e lattice at $\theta=43.6^\circ$ on an
  infinite cylinder geometry. The color scale shows the expectation
  value $\langle \hat{x} \rangle$ of the eigenstates, where $\hat{x}$
  denotes the position along the open direction of cylinder and its
  range is normalized to $[-1, 1]$. We choose $\mu=-4.25$, $h_z=0.1$
  for the upper panel $(a)$ and $\mu=-3.85$, $h_z=0.075$ for the lower
  panel $(b)$, which corresponds to the non-Abelian phases with $C=-1$
  and $C=-5$, respectively. Along the open $x$-direction, 50 Moir\'e unit
  cells are used in the calculations, and $\tilde{k}_y$ denotes the Moir\'e
  momentum along the $y$-direction.
  The other parameters are the
  same as those in Fig. \ref{fig:chern}(b).}
\label{fig:edge}
\end{figure}

\subsection{Numerical results at commensurate angle}

The results obtained from an effective continuum model as in previous
sections can be verified when the twisted angle is commensurate,
\textit{i.e.}, a Moir\'e lattice forms. To implement this, we look at
the special case of very short-range interlayer hopping: the in-plane
hopping distance $d_{ij}$ is smaller than $a_0/2$. In this case, the hopping between two
layers almost does not change the momentum, therefore,
$t_z(\kv_a,\kv_b)$ is well approximated by a constant $t_z^0$, which
can be calculated by averaging the amplitude of hopping inside a
Moir\'e unit cell. For $\theta=53.13^\circ$,
$t_z^0=g_0(1+4\times 0.8919)/5=0.9135g_0$, in which only the
nearest-neighboring and next-nearest-neighboring interlayer hopping
are counted. In a similar way, $t_z^0=0.6796g_0$ can be worked out for
$\theta=43.6^\circ$. We find that the energy bands of continuum model
with such values of $t_z^0$ are in good agreement with the bands
calculated from Moir\'e lattice.

We compute the Chern number using the gauge-independent method
developed in Ref. \cite{fu.ha.05}, and present in
Fig. \ref{fig:chern}(b) the phase diagram of Moir\'e lattice at
$\theta=43.6^\circ$ with short-range interlayer hopping. There are two
non-Abelian topological phases with $C=-1$ and $C=-5$ respectively,
which corresponds to one or three FSs enclosing the $\Gamma_0$
piont. The phase with $C=-5$ is specific to this bilayer system,
which does not exist in the single-layer case
\cite{sa.ta.10,ma.ya.22}. The numerical phase boundaries in
Fig. \ref{fig:chern}(b) and analytical results in
Fig. \ref{fig:chern}(a) match very well. Note that the non-Abelian
phase boundary in Eq. (\ref{eq:anal_boun}) is independent of twisted
angle $\theta$ at first glance, but it has a prerequisite that
time-reversal symmetry should be broken, which is determined by $\theta$.

The non-Abelian phases can be further confirmed by their edge states,
as demonstrated in Fig. \ref{fig:edge}. The twisted Rashba bilayer
shows a fully gapped bulk with one and five chiral edge modes,
corresponding to the non-Abelian phases with $C=-1$ and $C=-5$,
respectively.

\section{Twisted bilayer with $d$-wave and $s_{\pm}$-wave pairings}
\label{sec:d+is}

\begin{figure}
  \includegraphics[width=0.90\columnwidth]{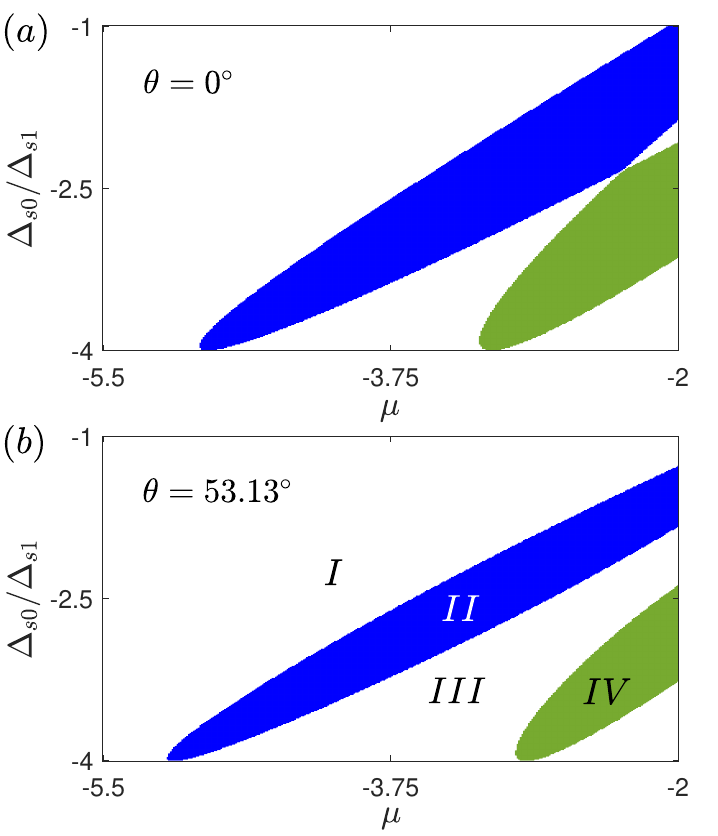}
  \caption{The phase diagram of second-order TSC in the plane of ratio
    $\Delta_{s0}/\Delta_{s1}$ and chemical potential $\mu$, for two
    twisted angles $\theta=0^\circ$ (a) and $\theta=53.13^\circ$
    (b). In order to make the nodal line of $s_{\pm}$-wave SC wind the
    $\Gamma_0$ point, the sign of $\Delta_{s0}$ and $\Delta_{s1}$ are
    chosen to be opposite. The blue (green) area denotes the second
    order TSC phase, in which the pairing nodes are enclosed by one
    (three) FS. The values of parameters in the calculations are
    $t=1,\lambda_0=0.25,g_0=0.25,\Delta_{d0}=0.5$.}
  \label{pd_sid}
\end{figure}

The hybrid Josephson junction, made up of extended
$s_{\pm}$-wave and $d$-wave SCs without a twist, is a second order topological SC \cite{zhu.19},
hosting MZMs at each corner of a sample. In this section, we
generalize this system by twisting two superconducting layers at an
angle. Due to the different pairing symmetries, the phase difference
between two layers is $\varphi=\frac{\pi}{2}$
\cite{ya.qi.18,ca.tu.21}, no matter what twisted angle is,
\textit{i.e.}, the time-reversal symmetry is always broken. The effective gap function in
Eq. (\ref{wDelta}) shows that the bilayer system is an odd parity SC
when $h_z=0$. Therefore, the topological properties are still
determined by the relation between SC pairing nodes and the FSs of normal
state \cite{yan.19,zhu.19,kh.ya.20}.

The effective pairing nodes for twisted
bilayer are given by
\begin{equation}
    \Delta_{a\pm
  b}(\mathbf{k}_a,\mathbf{k}_b)=\frac{1}{2}[\Delta_{d}(\mathbf{k}_a)\pm
        i\Delta_{s}(\mathbf{k}_b)]=0,
\end{equation}
that is, the nodes can be obtained by solving the equations
$\Delta_{d}(\mathbf{k}_a)=0$ and $\Delta_{s}(\mathbf{k}_b)=0$
simultaneously. There are totally four nodes in the BZ of unrotated
plane, and the coordinates $(Q_x,Q_y)$ of
the node in the first quadrant are
\begin{equation}\label{node}
  Q_x=(\cos\frac{\theta}{2}+\sin\frac{\theta}{2})K, \quad
  Q_y=(\cos\frac{\theta}{2}-\sin\frac{\theta}{2})K,
\end{equation}
in which $\sqrt{2} K$ ($K>0$) is the distance between $(Q_x,Q_y)$ and $(0,0)$ points,
and $K$ is determined by the following equation
\begin{equation}
    \cos(K\cos\theta )\cos(K\sin\theta) =-\frac{\Delta_{s0}}{4\Delta_{s1}}.
\end{equation}
The other three nodes are connected with the node $(Q_x,Q_y)$ by a $C_4$
rotation.
In contrast to the twisted $d$-wave bilayer in Sec. \ref{sec:d+id}
where the pairing nodes are fixed at the TRI points, the node in Eq.
(\ref{node}) is removable in the $\kv$-space, the so-called removable
Dirac pairing node (RDPN) \cite{yan.19}, depending on the ratio
$\Delta_{s0} / \Delta_{s1}$ as well as the twisted angle $\theta$.

The winding number $w$ around a RDPN can be calculated
\cite{yan.19,kh.ya.20},
\begin{equation}
w=\frac{1}{2\pi i}\oint \Delta_{odd}^{-1}\partial_k\Delta_{odd}dk
\end{equation}
with $\Delta_{odd}=\Delta_{a+b}\lambda(\kv)$, which turn out to be
$1, -1, 1, -1$ for the four pairing nodes of twisted bilayer,
respectively. When four pairing nodes are enclosed by a normal-state
FS, the net sum of winding numbers is zero, resulting in a trivial
first-order topological phase.  But if four nodes are enclosed by an
odd number of FSs, a pair of FSs cannot be continuously deformed to
annihilate with each other without crossing any RDPNs, and
consequently the system realizes a second-order TSC
\cite{yan.19,kh.ya.20}. By substituting the coordinates of pairing
node into the exact dispersion $E_m$ of normal state, the criterion
for the second-order TSC can be written as
\begin{equation}\label{sgnQ}
(-1)^N=\prod_{m=1}^{4}sgn[E_{m}(Q_{x},Q_{y})],
\end{equation}
where $N$ is the number of FSs enclosing the pairing nodes, and only
one node $(Q_{x},Q_{y})$ is used due to the $C_4$ rotation symmetry of
nodes.

\begin{table}
\begin{ruledtabular}
\begin{tabular}{cccc}
phase&N&$\prod_{m}sgn[E_{m}(Q_{x},Q_{y})]$&topology\\
\hline
(I)   &0&   1 & Trivial\\
(II)  &1&   -1 & Second order TSC\\
(III) &2&   1 & Trivial\\
(IV)  &3&   -1 & Second order TSC\\
\end{tabular}
\end{ruledtabular}
\caption{The topology is determined by the number of FSs that enclose the
  pairing nodes.}
\label{table}
\end{table}

\begin{figure*}[ht]
\includegraphics[width=1.95\columnwidth]{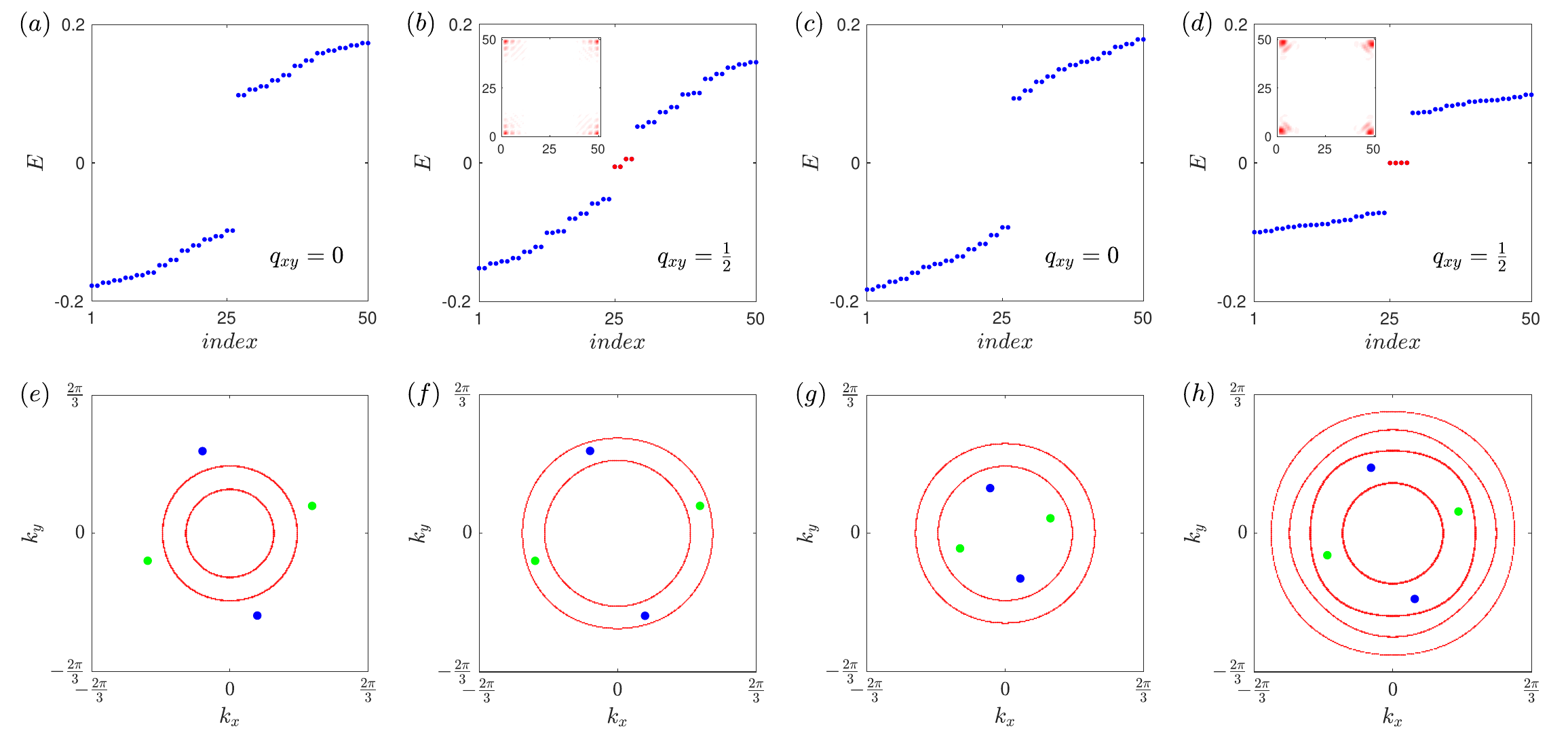}
\caption{The upper four panels $(a)-(d)$: the energy spectrum around
  zero energy for a $50\times50$ bilayer sample with a twisted angle
  $\theta=53.13^\circ$. The $50\times50$ sample is chosen according to
  the lattice number
  of $d$-wave layer (not the lattice number of Moir\'e lattice). The red dots in panels $(b)$ and $(d)$
  represent the Majorana corner modes, whose
  probability density profiles are shown in the insets of
  corresponding panels.  The lower four panels $(e)-(f)$: the pairing
  nodes and FSs of a bulk sample, in which the pairing nodes with
  winding number $w=1(w=-1)$ are denoted by the green (blue) dots.
  The values of $(\mu, \Delta_{s0} / \Delta_{s1})$ are chosen to be
  (-4.25, -2.5) in panels $(a)$ and $(e)$; (-3.3, -2.5) in panels
  $(b)$ and $(f)$; (-3.5, -3.5) in panels $(c)$ and $(g)$; (-2.2,
  -3.0) in panels $(d)$ and $(g)$. These four sets of parameters
  correspond to the points in four phases of Fig. \ref{pd_sid}(b) (from
  $I$ to $IV$), respectively. The values of other parameters are the
  same as those in Fig. \ref{pd_sid}(b).}
\label{corner}
\end{figure*}

The phase diagrams for second-order TSC are shown in Fig. \ref{pd_sid}
for twisted angle $\theta=0^\circ$ and $\theta=53.13^\circ$, which are
computed by adopting the criterion in Eq. (\ref{sgnQ}). We use the
long-range interlayer hopping $g_{ij}$ in Eq. (\ref{gij}), with a
cutoff of $d_{ij}$ being up to $\sqrt{2} a_0$.
The blue and green areas in the phase diagram denote
the second-order TSC phases, where the pairing nodes are enclosed by
one and three FSs, respectively; see table \ref{table} and
Fig. \ref{corner}. In the other areas, the pairing nodes are enclosed
by an even number of FSs and therefore the topology is trivial.
One can see that the regions of second order TSC in
Fig. \ref{pd_sid}(b) is smaller than those in Fig. \ref{pd_sid}(a).
The reason is as follows: twisting the bilayer increases the hopping
distance between two layers, which then decreases the averaged
interlayer coupling. This decrease in interlayer coupling changes the
distance between the two split FSs, making it "difficult" to enclose the
four Dirac pairing nodes by one or three FSs. Note that the positions
of paring nodes will also change with a nonzero twisted angle.
We do not address the region of small twist angles, as the Moir\'e
bands are highly complex in this regime, preventing us from accurately
analyzing the normal-state Fermi surface.

To support the above analysis of second order TSC, we 
diagonalize the real-space Hamiltonian of a finite sample and
show in Fig. \ref{corner} the energy spectrum near zero energy.
The bilayer sample is chosen to be a square containing $50\times50$
lattice sites of the $d$-wave layer. The twisted $s_\pm$-wave layer is
cut to match the same area as the $d$-wave layer. In
the second-order topological phases such as Fig. \ref{corner}(b) and
(d), there are four zero-energy modes whose wave functions are
localized at the four corners of the sample. However, no zero-energy
modes are observed for the topologically trivial phases; see
Fig. \ref{corner}(a) and (c). These corner modes are in good agreement
with the bulk topology, which is determined by the relative configurations of
pairing nodes and FSs, as illustrated in the lower four panels of Fig. \ref{corner}.

We further calculate quadrupole moments $q_{xy}$ using the real-space
formula \cite{ka.sh.19,wh.wa.19,li.fu.20,ya.li.21},
\begin{equation}\label{qxy}
q_{xy}=\frac{1}{2\pi}\textnormal{Im}\log \Big[
     \det(U^\dagger\hat{Q} U )\sqrt{\det(Q^\dagger)} \Big].
\end{equation}
In the above equation, $\hat{Q}=\exp[i2\pi\hat{x}\hat{y}/(L_xL_y)]$, where
$\hat{x}$($\hat{y}$) is the position operator along the $x$($y$)
direction and $L_x$ ($L_y$) is the corresponding system size; $U$ is a
matrix constructed by arranging the occupied eigenstates column by
column.
The calculated value of $q_{xy}$ using Eq. (\ref{qxy}) is
$\frac{1}{2}$ in the second-order topological phase (the blue and
green regions in Fig. \ref{pd_sid}), while it is zero in the
topologically trivial phase. The quantization of $q_{xy}$ to either 0
or $\frac{1}{2}$ is protected by the particle-hole symmetry \cite{li.fu.20}. The
nonzero value of $q_{xy}$ is consistent with the emergence of corner
states in a finite sample, as shown in Fig. \ref{corner}(b) and (d).

Due to the $C_4$ symmetry of the system, the FSs surround the four
pairing nodes at the same time. If the $C_4$ symmetry is broken, for
example, when the intra-layer hopping amplitudes along the $\hat{x}$
and $\hat{y}$ directions are different, or when both Dresselhaus and
Rashba SOCs are present, it is possible for the FSs to enclose only
two pairing nodes \cite{li.lu.23}. This will lead to a first-order topological phase
\cite{kh.ya.20}, but still with an even Chern number.

\section{Conclusions and Discussions}
\label{sec:conclusion}

In summary, we study the effect of Rashba SOC on the twisted bilayer
SC with spin-singlet pairings. Our analysis is in the framework of
mean-field theory, and we do not investigate the stability of SC
phase, but assume that the interlayer coupling is a minor perturbation
that does not change the magnitude of SC order parameter in each
layer \cite{ha.tu.22}.  The relative phase between two layers is
determined by minimizing the ground state energy.  Following Sato
\textit{et al.} \cite{sa.fu.10}, the even-parity bilayer SC with
Rashba SOC is transformed into an effective odd-parity SC, within the
approximation of circular FS. This is the starting point of our work.
For the case of twisted $d$-wave bilayer, we find two non-Abelian
topological phases with Chern number $C=-1$ and $C=-5$, and the phase
with $C=-5$ is specific to the bilayer which does not exist in the
single-layer $d+id$ SC.  The phase boundaries of non-Abelian
topological phases are obtained by counting the number of FSs that
cross the TRI $\Gamma_0$ point, which are in good agreement with
the numerical results at the twisted angle of Moir\'e lattice.  For
the twisted case when one layer is $d$-wave SC but the other layer is
$s_{\pm}$-wave SC, the second-order TSC phase appears when the pairing
nodes are enclosed by FSs for one or three times. We perform numerical
calculations for Moir\'e lattice at $53.13^\circ$, and show that the
regions of second-order TSC phase are narrowed when the twisted angle
is nonzero.

Experimentally, the nontrivial topological gap can be detected by
probing the local density of states using scanning tunneling
microscopy (STM) or by examining quasiparticle dispersion through
angle-resolved photoemission spectroscopy (ARPES).  Thermal transport
experiments offer a promising approach to identifying the values of
Chern number in topological superconductor phases \cite{sa.an.17}.  Each gapless
chiral edge mode contributes a quantized thermal Hall conductance \cite{se.ma.99}, and
the number of these modes is directly related to the Chern
number. Majorana corner states, localized at the corners of a
well-defined sample as discussed in Sec. \ref{sec:d+is}, can be
detected by using STM to measure the differential conductance. These
Majorana zero modes are expected to exhibit a characteristic zero-bias
conductance peak in the tunneling spectrum \cite{la.le.09}.


To experimentally realize the bilayer model studied in this paper, the
key step is to induce SOC in unconventional $d$-wave and $s_\pm$-wave
superconductors. This is experimentally feasible, considering the
great advances in van der Waals (vdW) stacking techniques.  Single
layer \cite{yu.ma.19} and twisted bilayer cuprate
\cite{zh.cu.23,zh.wa.23}, as well as single-layer iron-based
superconductors \cite{li.zh.12} have been successfully fabricated and
investigated.
A way to produce a
Rashba superconductor is to bring these materials into proximity with
a 2D material exhibiting large spin-orbit coupling. A promising
candidate is monolayer $WTe_2$, which has recently been confirmed as a
high-temperature topological insulator in experiments \cite{wu.fa.18}.
Another candidate for constructing cuprate-based heterostructures is
2D ultrathin semiconducting $Bi_2O_2Se$. This material has
excellent lattice constant matching and a similar composition to
Bi-based cuprate, resulting in strong coupling between them
\cite{ma.ya.22,wu.yu.17}.
Artificial superlattices of $CeCoIn_5$
are also a promising platform for our theoretical model, featuring $d$-wave pairing and strong
Rashba SOC \cite{na.te.21}. The Rashba SOC is introduced by
breaking the inversion symmetry locally (bicolor stacking)
\cite{mi.sh.11} or globally (tricolor stacking) \cite{na.is.17}, and
is tunable by adjusting the layer thickness.  After preparing such
artificial heterostructures, one can stack these structures together
and assemble them at a relative angle. While this presents an
experimental challenge, it may be possible in the future.

\begin{acknowledgments}
  Stimulating discussions with  Chuanshu Xu, Zhesen Yang, Chuanshuai Huang, and Zhenghao
  Yang are gratefully acknowledged. This work is supported by
  the National Natural Science Foundation of China (Grant
  No. 11974293).
\end{acknowledgments}


\begin{thebibliography}{67}%
\makeatletter
\providecommand \@ifxundefined [1]{%
 \@ifx{#1\undefined}
}%
\providecommand \@ifnum [1]{%
 \ifnum #1\expandafter \@firstoftwo
 \else \expandafter \@secondoftwo
 \fi
}%
\providecommand \@ifx [1]{%
 \ifx #1\expandafter \@firstoftwo
 \else \expandafter \@secondoftwo
 \fi
}%
\providecommand \natexlab [1]{#1}%
\providecommand \enquote  [1]{``#1''}%
\providecommand \bibnamefont  [1]{#1}%
\providecommand \bibfnamefont [1]{#1}%
\providecommand \citenamefont [1]{#1}%
\providecommand \href@noop [0]{\@secondoftwo}%
\providecommand \href [0]{\begingroup \@sanitize@url \@href}%
\providecommand \@href[1]{\@@startlink{#1}\@@href}%
\providecommand \@@href[1]{\endgroup#1\@@endlink}%
\providecommand \@sanitize@url [0]{\catcode `\\12\catcode `\$12\catcode
  `\&12\catcode `\#12\catcode `\^12\catcode `\_12\catcode `\%12\relax}%
\providecommand \@@startlink[1]{}%
\providecommand \@@endlink[0]{}%
\providecommand \url  [0]{\begingroup\@sanitize@url \@url }%
\providecommand \@url [1]{\endgroup\@href {#1}{\urlprefix }}%
\providecommand \urlprefix  [0]{URL }%
\providecommand \Eprint [0]{\href }%
\providecommand \doibase [0]{https://doi.org/}%
\providecommand \selectlanguage [0]{\@gobble}%
\providecommand \bibinfo  [0]{\@secondoftwo}%
\providecommand \bibfield  [0]{\@secondoftwo}%
\providecommand \translation [1]{[#1]}%
\providecommand \BibitemOpen [0]{}%
\providecommand \bibitemStop [0]{}%
\providecommand \bibitemNoStop [0]{.\EOS\space}%
\providecommand \EOS [0]{\spacefactor3000\relax}%
\providecommand \BibitemShut  [1]{\csname bibitem#1\endcsname}%
\let\auto@bib@innerbib\@empty
\bibitem [{\citenamefont {Qi}\ and\ \citenamefont {Zhang}(2011)}]{qi.zh.11}%
  \BibitemOpen
  \bibfield  {author} {\bibinfo {author} {\bibfnamefont {X.-L.}\ \bibnamefont
  {Qi}}\ and\ \bibinfo {author} {\bibfnamefont {S.-C.}\ \bibnamefont {Zhang}},\
  }\bibfield  {title} {\bibinfo {title} {Topological insulators and
  superconductors},\ }\href {https://doi.org/10.1103/RevModPhys.83.1057}
  {\bibfield  {journal} {\bibinfo  {journal} {Rev. Mod. Phys.}\ }\textbf
  {\bibinfo {volume} {83}},\ \bibinfo {pages} {1057} (\bibinfo {year}
  {2011})}\BibitemShut {NoStop}%
\bibitem [{\citenamefont {Sato}\ and\ \citenamefont {Ando}(2017)}]{sa.an.17}%
  \BibitemOpen
  \bibfield  {author} {\bibinfo {author} {\bibfnamefont {M.}~\bibnamefont
  {Sato}}\ and\ \bibinfo {author} {\bibfnamefont {Y.}~\bibnamefont {Ando}},\
  }\bibfield  {title} {\bibinfo {title} {Topological superconductors: a
  review},\ }\href {http://stacks.iop.org/0034-4885/80/i=7/a=076501} {\bibfield
   {journal} {\bibinfo  {journal} {Reports on Progress in Physics}\ }\textbf
  {\bibinfo {volume} {80}},\ \bibinfo {pages} {076501} (\bibinfo {year}
  {2017})}\BibitemShut {NoStop}%
\bibitem [{\citenamefont {Read}\ and\ \citenamefont {Green}(2000)}]{re.gr.00}%
  \BibitemOpen
  \bibfield  {author} {\bibinfo {author} {\bibfnamefont {N.}~\bibnamefont
  {Read}}\ and\ \bibinfo {author} {\bibfnamefont {D.}~\bibnamefont {Green}},\
  }\bibfield  {title} {\bibinfo {title} {Paired states of fermions in two
  dimensions with breaking of parity and time-reversal symmetries and the
  fractional quantum hall effect},\ }\href
  {https://doi.org/10.1103/PhysRevB.61.10267} {\bibfield  {journal} {\bibinfo
  {journal} {Phys. Rev. B}\ }\textbf {\bibinfo {volume} {61}},\ \bibinfo
  {pages} {10267} (\bibinfo {year} {2000})}\BibitemShut {NoStop}%
\bibitem [{\citenamefont {Kitaev}(2001)}]{kita.01}%
  \BibitemOpen
  \bibfield  {author} {\bibinfo {author} {\bibfnamefont {A.~Y.}\ \bibnamefont
  {Kitaev}},\ }\bibfield  {title} {\bibinfo {title} {Unpaired majorana fermions
  in quantum wires},\ }\href {http://stacks.iop.org/1063-7869/44/i=10S/a=S29}
  {\bibfield  {journal} {\bibinfo  {journal} {Physics-Uspekhi}\ }\textbf
  {\bibinfo {volume} {44}},\ \bibinfo {pages} {131} (\bibinfo {year}
  {2001})}\BibitemShut {NoStop}%
\bibitem [{\citenamefont {Ivanov}(2001)}]{ivan.01}%
  \BibitemOpen
  \bibfield  {author} {\bibinfo {author} {\bibfnamefont {D.~A.}\ \bibnamefont
  {Ivanov}},\ }\bibfield  {title} {\bibinfo {title} {Non-abelian statistics of
  half-quantum vortices in $\mathit{p}$-wave superconductors},\ }\href
  {https://doi.org/10.1103/PhysRevLett.86.268} {\bibfield  {journal} {\bibinfo
  {journal} {Phys. Rev. Lett.}\ }\textbf {\bibinfo {volume} {86}},\ \bibinfo
  {pages} {268} (\bibinfo {year} {2001})}\BibitemShut {NoStop}%
\bibitem [{\citenamefont {Fu}\ and\ \citenamefont {Kane}(2008)}]{fu.ka.08}%
  \BibitemOpen
  \bibfield  {author} {\bibinfo {author} {\bibfnamefont {L.}~\bibnamefont
  {Fu}}\ and\ \bibinfo {author} {\bibfnamefont {C.~L.}\ \bibnamefont {Kane}},\
  }\bibfield  {title} {\bibinfo {title} {Superconducting proximity effect and
  majorana fermions at the surface of a topological insulator},\ }\href
  {https://doi.org/10.1103/PhysRevLett.100.096407} {\bibfield  {journal}
  {\bibinfo  {journal} {Phys. Rev. Lett.}\ }\textbf {\bibinfo {volume} {100}},\
  \bibinfo {pages} {096407} (\bibinfo {year} {2008})}\BibitemShut {NoStop}%
\bibitem [{\citenamefont {Sau}\ \emph {et~al.}(2010)\citenamefont {Sau},
  \citenamefont {Lutchyn}, \citenamefont {Tewari},\ and\ \citenamefont
  {Das~Sarma}}]{sa.lu.10}%
  \BibitemOpen
  \bibfield  {author} {\bibinfo {author} {\bibfnamefont {J.~D.}\ \bibnamefont
  {Sau}}, \bibinfo {author} {\bibfnamefont {R.~M.}\ \bibnamefont {Lutchyn}},
  \bibinfo {author} {\bibfnamefont {S.}~\bibnamefont {Tewari}},\ and\ \bibinfo
  {author} {\bibfnamefont {S.}~\bibnamefont {Das~Sarma}},\ }\bibfield  {title}
  {\bibinfo {title} {Generic new platform for topological quantum computation
  using semiconductor heterostructures},\ }\href
  {https://doi.org/10.1103/PhysRevLett.104.040502} {\bibfield  {journal}
  {\bibinfo  {journal} {Phys. Rev. Lett.}\ }\textbf {\bibinfo {volume} {104}},\
  \bibinfo {pages} {040502} (\bibinfo {year} {2010})}\BibitemShut {NoStop}%
\bibitem [{\citenamefont {Oreg}\ \emph {et~al.}(2010)\citenamefont {Oreg},
  \citenamefont {Refael},\ and\ \citenamefont {von Oppen}}]{or.re.10}%
  \BibitemOpen
  \bibfield  {author} {\bibinfo {author} {\bibfnamefont {Y.}~\bibnamefont
  {Oreg}}, \bibinfo {author} {\bibfnamefont {G.}~\bibnamefont {Refael}},\ and\
  \bibinfo {author} {\bibfnamefont {F.}~\bibnamefont {von Oppen}},\ }\bibfield
  {title} {\bibinfo {title} {Helical liquids and majorana bound states in
  quantum wires},\ }\href {https://doi.org/10.1103/PhysRevLett.105.177002}
  {\bibfield  {journal} {\bibinfo  {journal} {Phys. Rev. Lett.}\ }\textbf
  {\bibinfo {volume} {105}},\ \bibinfo {pages} {177002} (\bibinfo {year}
  {2010})}\BibitemShut {NoStop}%
\bibitem [{\citenamefont {Alicea}(2010)}]{alic.10}%
  \BibitemOpen
  \bibfield  {author} {\bibinfo {author} {\bibfnamefont {J.}~\bibnamefont
  {Alicea}},\ }\bibfield  {title} {\bibinfo {title} {Majorana fermions in a
  tunable semiconductor device},\ }\href
  {https://doi.org/10.1103/PhysRevB.81.125318} {\bibfield  {journal} {\bibinfo
  {journal} {Phys. Rev. B}\ }\textbf {\bibinfo {volume} {81}},\ \bibinfo
  {pages} {125318} (\bibinfo {year} {2010})}\BibitemShut {NoStop}%
\bibitem [{\citenamefont {Alicea}(2012)}]{alic.12}%
  \BibitemOpen
  \bibfield  {author} {\bibinfo {author} {\bibfnamefont {J.}~\bibnamefont
  {Alicea}},\ }\bibfield  {title} {\bibinfo {title} {New directions in the
  pursuit of majorana fermions in solid state systems},\ }\href
  {https://doi.org/10.1088/0034-4885/75/7/076501} {\bibfield  {journal}
  {\bibinfo  {journal} {Reports on Progress in Physics}\ }\textbf {\bibinfo
  {volume} {75}},\ \bibinfo {pages} {076501} (\bibinfo {year}
  {2012})}\BibitemShut {NoStop}%
\bibitem [{\citenamefont {Yu}\ \emph {et~al.}(2019)\citenamefont {Yu},
  \citenamefont {Ma}, \citenamefont {Cai}, \citenamefont {Zhong}, \citenamefont
  {Ye}, \citenamefont {Shen}, \citenamefont {Gu}, \citenamefont {Chen},\ and\
  \citenamefont {Zhang}}]{yu.ma.19}%
  \BibitemOpen
  \bibfield  {author} {\bibinfo {author} {\bibfnamefont {Y.}~\bibnamefont
  {Yu}}, \bibinfo {author} {\bibfnamefont {L.}~\bibnamefont {Ma}}, \bibinfo
  {author} {\bibfnamefont {P.}~\bibnamefont {Cai}}, \bibinfo {author}
  {\bibfnamefont {R.}~\bibnamefont {Zhong}}, \bibinfo {author} {\bibfnamefont
  {C.}~\bibnamefont {Ye}}, \bibinfo {author} {\bibfnamefont {J.}~\bibnamefont
  {Shen}}, \bibinfo {author} {\bibfnamefont {G.~D.}\ \bibnamefont {Gu}},
  \bibinfo {author} {\bibfnamefont {X.~H.}\ \bibnamefont {Chen}},\ and\
  \bibinfo {author} {\bibfnamefont {Y.}~\bibnamefont {Zhang}},\ }\bibfield
  {title} {\bibinfo {title} {High-temperature superconductivity in monolayer
  $Bi_2Sr_2CaCu_2O_{8+\delta}$},\ }\href {https://doi.org/10.1038/s41586-019-1718-x}
  {\bibfield  {journal} {\bibinfo  {journal} {Nature}\ }\textbf {\bibinfo
  {volume} {575}},\ \bibinfo {pages} {156} (\bibinfo {year}
  {2019})}\BibitemShut {NoStop}%
\bibitem [{\citenamefont {Zhao}\ \emph {et~al.}(2019)\citenamefont {Zhao},
  \citenamefont {Poccia}, \citenamefont {Panetta}, \citenamefont {Yu},
  \citenamefont {Johnson}, \citenamefont {Yoo}, \citenamefont {Zhong},
  \citenamefont {Gu}, \citenamefont {Watanabe}, \citenamefont {Taniguchi},
  \citenamefont {Postolova}, \citenamefont {Vinokur},\ and\ \citenamefont
  {Kim}}]{zh.po.19}%
  \BibitemOpen
  \bibfield  {author} {\bibinfo {author} {\bibfnamefont {S.~Y.~F.}\
  \bibnamefont {Zhao}}, \bibinfo {author} {\bibfnamefont {N.}~\bibnamefont
  {Poccia}}, \bibinfo {author} {\bibfnamefont {M.~G.}\ \bibnamefont {Panetta}},
  \bibinfo {author} {\bibfnamefont {C.}~\bibnamefont {Yu}}, \bibinfo {author}
  {\bibfnamefont {J.~W.}\ \bibnamefont {Johnson}}, \bibinfo {author}
  {\bibfnamefont {H.}~\bibnamefont {Yoo}}, \bibinfo {author} {\bibfnamefont
  {R.}~\bibnamefont {Zhong}}, \bibinfo {author} {\bibfnamefont {G.~D.}\
  \bibnamefont {Gu}}, \bibinfo {author} {\bibfnamefont {K.}~\bibnamefont
  {Watanabe}}, \bibinfo {author} {\bibfnamefont {T.}~\bibnamefont {Taniguchi}},
  \bibinfo {author} {\bibfnamefont {S.~V.}\ \bibnamefont {Postolova}}, \bibinfo
  {author} {\bibfnamefont {V.~M.}\ \bibnamefont {Vinokur}},\ and\ \bibinfo
  {author} {\bibfnamefont {P.}~\bibnamefont {Kim}},\ }\bibfield  {title}
  {\bibinfo {title} {Sign-reversing hall effect in atomically thin
  high-temperature
  ${\mathrm{bi}}_{2.1}{\mathrm{sr}}_{1.9}{\mathrm{cacu}}_{2.0}{\mathrm{o}}_{8+\ensuremath{\delta}}$
  superconductors},\ }\href {https://doi.org/10.1103/PhysRevLett.122.247001}
  {\bibfield  {journal} {\bibinfo  {journal} {Phys. Rev. Lett.}\ }\textbf
  {\bibinfo {volume} {122}},\ \bibinfo {pages} {247001} (\bibinfo {year}
  {2019})}\BibitemShut {NoStop}%
\bibitem [{\citenamefont {Bistritzer}\ and\ \citenamefont
  {MacDonald}(2011)}]{bi.ma.11}%
  \BibitemOpen
  \bibfield  {author} {\bibinfo {author} {\bibfnamefont {R.}~\bibnamefont
  {Bistritzer}}\ and\ \bibinfo {author} {\bibfnamefont {A.~H.}\ \bibnamefont
  {MacDonald}},\ }\bibfield  {title} {\bibinfo {title} {Moir{\'e} bands in
  twisted double-layer graphene},\ }\href
  {https://doi.org/10.1073/pnas.1108174108} {\bibfield  {journal} {\bibinfo
  {journal} {Proceedings of the National Academy of Sciences}\ }\textbf
  {\bibinfo {volume} {108}},\ \bibinfo {pages} {12233} (\bibinfo {year}
  {2011})},\ \Eprint
  {https://arxiv.org/abs/https://www.pnas.org/content/108/30/12233.full.pdf}
  {https://www.pnas.org/content/108/30/12233.full.pdf} \BibitemShut {NoStop}%
\bibitem [{\citenamefont {Cao}\ \emph {et~al.}(2018{\natexlab{a}})\citenamefont
  {Cao}, \citenamefont {Fatemi}, \citenamefont {Fang}, \citenamefont
  {Watanabe}, \citenamefont {Taniguchi}, \citenamefont {Kaxiras},\ and\
  \citenamefont {Jarillo-Herrero}}]{ca.fa.18a}%
  \BibitemOpen
  \bibfield  {author} {\bibinfo {author} {\bibfnamefont {Y.}~\bibnamefont
  {Cao}}, \bibinfo {author} {\bibfnamefont {V.}~\bibnamefont {Fatemi}},
  \bibinfo {author} {\bibfnamefont {S.}~\bibnamefont {Fang}}, \bibinfo {author}
  {\bibfnamefont {K.}~\bibnamefont {Watanabe}}, \bibinfo {author}
  {\bibfnamefont {T.}~\bibnamefont {Taniguchi}}, \bibinfo {author}
  {\bibfnamefont {E.}~\bibnamefont {Kaxiras}},\ and\ \bibinfo {author}
  {\bibfnamefont {P.}~\bibnamefont {Jarillo-Herrero}},\ }\bibfield  {title}
  {\bibinfo {title} {Unconventional superconductivity in magic-angle graphene
  superlattices},\ }\href {https://doi.org/10.1038/nature26160} {\bibfield
  {journal} {\bibinfo  {journal} {Nature}\ }\textbf {\bibinfo {volume} {556}},\
  \bibinfo {pages} {43} (\bibinfo {year} {2018}{\natexlab{a}})}\BibitemShut
  {NoStop}%
\bibitem [{\citenamefont {Cao}\ \emph {et~al.}(2018{\natexlab{b}})\citenamefont
  {Cao}, \citenamefont {Fatemi}, \citenamefont {Demir}, \citenamefont {Fang},
  \citenamefont {Tomarken}, \citenamefont {Luo}, \citenamefont
  {Sanchez-Yamagishi}, \citenamefont {Watanabe}, \citenamefont {Taniguchi},
  \citenamefont {Kaxiras}, \citenamefont {Ashoori},\ and\ \citenamefont
  {Jarillo-Herrero}}]{ca.fa.18b}%
  \BibitemOpen
  \bibfield  {author} {\bibinfo {author} {\bibfnamefont {Y.}~\bibnamefont
  {Cao}}, \bibinfo {author} {\bibfnamefont {V.}~\bibnamefont {Fatemi}},
  \bibinfo {author} {\bibfnamefont {A.}~\bibnamefont {Demir}}, \bibinfo
  {author} {\bibfnamefont {S.}~\bibnamefont {Fang}}, \bibinfo {author}
  {\bibfnamefont {S.~L.}\ \bibnamefont {Tomarken}}, \bibinfo {author}
  {\bibfnamefont {J.~Y.}\ \bibnamefont {Luo}}, \bibinfo {author} {\bibfnamefont
  {J.~D.}\ \bibnamefont {Sanchez-Yamagishi}}, \bibinfo {author} {\bibfnamefont
  {K.}~\bibnamefont {Watanabe}}, \bibinfo {author} {\bibfnamefont
  {T.}~\bibnamefont {Taniguchi}}, \bibinfo {author} {\bibfnamefont
  {E.}~\bibnamefont {Kaxiras}}, \bibinfo {author} {\bibfnamefont {R.~C.}\
  \bibnamefont {Ashoori}},\ and\ \bibinfo {author} {\bibfnamefont
  {P.}~\bibnamefont {Jarillo-Herrero}},\ }\bibfield  {title} {\bibinfo {title}
  {Correlated insulator behaviour at half-filling in magic-angle graphene
  superlattices},\ }\href {https://doi.org/10.1038/nature26154} {\bibfield
  {journal} {\bibinfo  {journal} {Nature}\ }\textbf {\bibinfo {volume} {556}},\
  \bibinfo {pages} {80} (\bibinfo {year} {2018}{\natexlab{b}})}\BibitemShut
  {NoStop}%
\bibitem [{\citenamefont {Wu}\ \emph {et~al.}(2019)\citenamefont {Wu},
  \citenamefont {Lovorn}, \citenamefont {Tutuc}, \citenamefont {Martin},\ and\
  \citenamefont {MacDonald}}]{wu.lo.19}%
  \BibitemOpen
  \bibfield  {author} {\bibinfo {author} {\bibfnamefont {F.}~\bibnamefont
  {Wu}}, \bibinfo {author} {\bibfnamefont {T.}~\bibnamefont {Lovorn}}, \bibinfo
  {author} {\bibfnamefont {E.}~\bibnamefont {Tutuc}}, \bibinfo {author}
  {\bibfnamefont {I.}~\bibnamefont {Martin}},\ and\ \bibinfo {author}
  {\bibfnamefont {A.~H.}\ \bibnamefont {MacDonald}},\ }\bibfield  {title}
  {\bibinfo {title} {Topological insulators in twisted transition metal
  dichalcogenide homobilayers},\ }\href
  {https://doi.org/10.1103/PhysRevLett.122.086402} {\bibfield  {journal}
  {\bibinfo  {journal} {Phys. Rev. Lett.}\ }\textbf {\bibinfo {volume} {122}},\
  \bibinfo {pages} {086402} (\bibinfo {year} {2019})}\BibitemShut {NoStop}%
\bibitem [{\citenamefont {Andrei}\ and\ \citenamefont
  {MacDonald}(2020)}]{an.ma.20}%
  \BibitemOpen
  \bibfield  {author} {\bibinfo {author} {\bibfnamefont {E.~Y.}\ \bibnamefont
  {Andrei}}\ and\ \bibinfo {author} {\bibfnamefont {A.~H.}\ \bibnamefont
  {MacDonald}},\ }\bibfield  {title} {\bibinfo {title} {Graphene bilayers with
  a twist},\ }\href {https://doi.org/10.1038/s41563-020-00840-0} {\bibfield
  {journal} {\bibinfo  {journal} {Nature Materials}\ }\textbf {\bibinfo
  {volume} {19}},\ \bibinfo {pages} {1265} (\bibinfo {year}
  {2020})}\BibitemShut {NoStop}%
\bibitem [{\citenamefont {Can}\ \emph {et~al.}(2021)\citenamefont {Can},
  \citenamefont {Tummuru}, \citenamefont {Day}, \citenamefont {Elfimov},
  \citenamefont {Damascelli},\ and\ \citenamefont {Franz}}]{ca.tu.21}%
  \BibitemOpen
  \bibfield  {author} {\bibinfo {author} {\bibfnamefont {O.}~\bibnamefont
  {Can}}, \bibinfo {author} {\bibfnamefont {T.}~\bibnamefont {Tummuru}},
  \bibinfo {author} {\bibfnamefont {R.~P.}\ \bibnamefont {Day}}, \bibinfo
  {author} {\bibfnamefont {I.}~\bibnamefont {Elfimov}}, \bibinfo {author}
  {\bibfnamefont {A.}~\bibnamefont {Damascelli}},\ and\ \bibinfo {author}
  {\bibfnamefont {M.}~\bibnamefont {Franz}},\ }\bibfield  {title} {\bibinfo
  {title} {High-temperature topological superconductivity in twisted
  double-layer copper oxides},\ }\href
  {https://doi.org/10.1038/s41567-020-01142-7} {\bibfield  {journal} {\bibinfo
  {journal} {Nature Physics}\ }\textbf {\bibinfo {volume} {17}},\ \bibinfo
  {pages} {519} (\bibinfo {year} {2021})}\BibitemShut {NoStop}%
\bibitem [{\citenamefont {Golubov}\ \emph {et~al.}(2004)\citenamefont
  {Golubov}, \citenamefont {Kupriyanov},\ and\ \citenamefont
  {Il'ichev}}]{go.ku.04}%
  \BibitemOpen
  \bibfield  {author} {\bibinfo {author} {\bibfnamefont {A.~A.}\ \bibnamefont
  {Golubov}}, \bibinfo {author} {\bibfnamefont {M.~Y.}\ \bibnamefont
  {Kupriyanov}},\ and\ \bibinfo {author} {\bibfnamefont {E.}~\bibnamefont
  {Il'ichev}},\ }\bibfield  {title} {\bibinfo {title} {The current-phase
  relation in josephson junctions},\ }\href
  {https://doi.org/10.1103/RevModPhys.76.411} {\bibfield  {journal} {\bibinfo
  {journal} {Rev. Mod. Phys.}\ }\textbf {\bibinfo {volume} {76}},\ \bibinfo
  {pages} {411} (\bibinfo {year} {2004})}\BibitemShut {NoStop}%
\bibitem [{\citenamefont {Yang}\ \emph {et~al.}(2018)\citenamefont {Yang},
  \citenamefont {Qin}, \citenamefont {Zhang}, \citenamefont {Fang},\ and\
  \citenamefont {Hu}}]{ya.qi.18}%
  \BibitemOpen
  \bibfield  {author} {\bibinfo {author} {\bibfnamefont {Z.}~\bibnamefont
  {Yang}}, \bibinfo {author} {\bibfnamefont {S.}~\bibnamefont {Qin}}, \bibinfo
  {author} {\bibfnamefont {Q.}~\bibnamefont {Zhang}}, \bibinfo {author}
  {\bibfnamefont {C.}~\bibnamefont {Fang}},\ and\ \bibinfo {author}
  {\bibfnamefont {J.}~\bibnamefont {Hu}},\ }\bibfield  {title} {\bibinfo
  {title} {$\ensuremath{\pi}$/2-josephson junction as a topological
  superconductor},\ }\href {https://doi.org/10.1103/PhysRevB.98.104515}
  {\bibfield  {journal} {\bibinfo  {journal} {Phys. Rev. B}\ }\textbf {\bibinfo
  {volume} {98}},\ \bibinfo {pages} {104515} (\bibinfo {year}
  {2018})}\BibitemShut {NoStop}%
\bibitem [{\citenamefont {Zhao}\ \emph {et~al.}(2023)\citenamefont {Zhao},
  \citenamefont {Cui}, \citenamefont {Volkov}, \citenamefont {Yoo},
  \citenamefont {Lee}, \citenamefont {Gardener}, \citenamefont {Akey},
  \citenamefont {Engelke}, \citenamefont {Ronen}, \citenamefont {Zhong},
  \citenamefont {Gu}, \citenamefont {Plugge}, \citenamefont {Tummuru},
  \citenamefont {Kim}, \citenamefont {Franz}, \citenamefont {Pixley},
  \citenamefont {Poccia},\ and\ \citenamefont {Kim}}]{zh.cu.23}%
  \BibitemOpen
  \bibfield  {author} {\bibinfo {author} {\bibfnamefont {S.~Y.~F.}\
  \bibnamefont {Zhao}}, \bibinfo {author} {\bibfnamefont {X.}~\bibnamefont
  {Cui}}, \bibinfo {author} {\bibfnamefont {P.~A.}\ \bibnamefont {Volkov}},
  \bibinfo {author} {\bibfnamefont {H.}~\bibnamefont {Yoo}}, \bibinfo {author}
  {\bibfnamefont {S.}~\bibnamefont {Lee}}, \bibinfo {author} {\bibfnamefont
  {J.~A.}\ \bibnamefont {Gardener}}, \bibinfo {author} {\bibfnamefont {A.~J.}\
  \bibnamefont {Akey}}, \bibinfo {author} {\bibfnamefont {R.}~\bibnamefont
  {Engelke}}, \bibinfo {author} {\bibfnamefont {Y.}~\bibnamefont {Ronen}},
  \bibinfo {author} {\bibfnamefont {R.}~\bibnamefont {Zhong}}, \bibinfo
  {author} {\bibfnamefont {G.}~\bibnamefont {Gu}}, \bibinfo {author}
  {\bibfnamefont {S.}~\bibnamefont {Plugge}}, \bibinfo {author} {\bibfnamefont
  {T.}~\bibnamefont {Tummuru}}, \bibinfo {author} {\bibfnamefont
  {M.}~\bibnamefont {Kim}}, \bibinfo {author} {\bibfnamefont {M.}~\bibnamefont
  {Franz}}, \bibinfo {author} {\bibfnamefont {J.~H.}\ \bibnamefont {Pixley}},
  \bibinfo {author} {\bibfnamefont {N.}~\bibnamefont {Poccia}},\ and\ \bibinfo
  {author} {\bibfnamefont {P.}~\bibnamefont {Kim}},\ }\bibfield  {title}
  {\bibinfo {title} {Time-reversal symmetry breaking superconductivity between
  twisted cuprate superconductors},\ }\href
  {https://doi.org/10.1126/science.abl8371} {\bibfield  {journal} {\bibinfo
  {journal} {Science}\ }\textbf {\bibinfo {volume} {382}},\ \bibinfo {pages}
  {1422} (\bibinfo {year} {2023})},\ \Eprint
  {https://arxiv.org/abs/https://www.science.org/doi/pdf/10.1126/science.abl8371}
  {https://www.science.org/doi/pdf/10.1126/science.abl8371} \BibitemShut
  {NoStop}%
\bibitem [{\citenamefont {Lee}\ \emph {et~al.}(2021)\citenamefont {Lee},
  \citenamefont {Lee}, \citenamefont {Kim}, \citenamefont {Choi}, \citenamefont
  {Park}, \citenamefont {Jang}, \citenamefont {Gu}, \citenamefont {Choi},
  \citenamefont {Cho}, \citenamefont {Lee},\ and\ \citenamefont
  {Lee}}]{le.le.21}%
  \BibitemOpen
  \bibfield  {author} {\bibinfo {author} {\bibfnamefont {J.}~\bibnamefont
  {Lee}}, \bibinfo {author} {\bibfnamefont {W.}~\bibnamefont {Lee}}, \bibinfo
  {author} {\bibfnamefont {G.-Y.}\ \bibnamefont {Kim}}, \bibinfo {author}
  {\bibfnamefont {Y.-B.}\ \bibnamefont {Choi}}, \bibinfo {author}
  {\bibfnamefont {J.}~\bibnamefont {Park}}, \bibinfo {author} {\bibfnamefont
  {S.}~\bibnamefont {Jang}}, \bibinfo {author} {\bibfnamefont {G.}~\bibnamefont
  {Gu}}, \bibinfo {author} {\bibfnamefont {S.-Y.}\ \bibnamefont {Choi}},
  \bibinfo {author} {\bibfnamefont {G.~Y.}\ \bibnamefont {Cho}}, \bibinfo
  {author} {\bibfnamefont {G.-H.}\ \bibnamefont {Lee}},\ and\ \bibinfo {author}
  {\bibfnamefont {H.-J.}\ \bibnamefont {Lee}},\ }\bibfield  {title} {\bibinfo
  {title} {Twisted van der waals josephson junction based on a high-tc
  superconductor},\ }\href {https://doi.org/10.1021/acs.nanolett.1c03906}
  {\bibfield  {journal} {\bibinfo  {journal} {Nano Letters}\ }\textbf {\bibinfo
  {volume} {21}},\ \bibinfo {pages} {10469} (\bibinfo {year} {2021})},\
  \bibinfo {note} {pMID: 34881903},\ \Eprint
  {https://arxiv.org/abs/https://doi.org/10.1021/acs.nanolett.1c03906}
  {https://doi.org/10.1021/acs.nanolett.1c03906} \BibitemShut {NoStop}%
\bibitem [{\citenamefont {Zhu}\ \emph {et~al.}(2023)\citenamefont {Zhu},
  \citenamefont {Wang}, \citenamefont {Wang}, \citenamefont {Hu}, \citenamefont
  {Gu}, \citenamefont {Zhu}, \citenamefont {Zhang},\ and\ \citenamefont
  {Xue}}]{zh.wa.23}%
  \BibitemOpen
  \bibfield  {author} {\bibinfo {author} {\bibfnamefont {Y.}~\bibnamefont
  {Zhu}}, \bibinfo {author} {\bibfnamefont {H.}~\bibnamefont {Wang}}, \bibinfo
  {author} {\bibfnamefont {Z.}~\bibnamefont {Wang}}, \bibinfo {author}
  {\bibfnamefont {S.}~\bibnamefont {Hu}}, \bibinfo {author} {\bibfnamefont
  {G.}~\bibnamefont {Gu}}, \bibinfo {author} {\bibfnamefont {J.}~\bibnamefont
  {Zhu}}, \bibinfo {author} {\bibfnamefont {D.}~\bibnamefont {Zhang}},\ and\
  \bibinfo {author} {\bibfnamefont {Q.-K.}\ \bibnamefont {Xue}},\ }\bibfield
  {title} {\bibinfo {title} {Persistent josephson tunneling between
  ${\mathrm{bi}}_{2}{\mathrm{sr}}_{2}{\mathrm{cacu}}_{2}{\mathrm{o}}_{8+x}$
  flakes twisted by ${45}^{\ensuremath{\circ}}$ across the superconducting
  dome},\ }\href {https://doi.org/10.1103/PhysRevB.108.174508} {\bibfield
  {journal} {\bibinfo  {journal} {Phys. Rev. B}\ }\textbf {\bibinfo {volume}
  {108}},\ \bibinfo {pages} {174508} (\bibinfo {year} {2023})}\BibitemShut
  {NoStop}%
\bibitem [{\citenamefont {Wang}\ \emph {et~al.}(2023)\citenamefont {Wang},
  \citenamefont {Zhu}, \citenamefont {Bai}, \citenamefont {Wang}, \citenamefont
  {Hu}, \citenamefont {Xie}, \citenamefont {Hu}, \citenamefont {Cui},
  \citenamefont {Huang}, \citenamefont {Chen}, \citenamefont {Ding},
  \citenamefont {Zhao}, \citenamefont {Li}, \citenamefont {Zhang},
  \citenamefont {Gu}, \citenamefont {Zhou}, \citenamefont {Zhu}, \citenamefont
  {Zhang},\ and\ \citenamefont {Xue}}]{wa.zh.23}%
  \BibitemOpen
  \bibfield  {author} {\bibinfo {author} {\bibfnamefont {H.}~\bibnamefont
  {Wang}}, \bibinfo {author} {\bibfnamefont {Y.}~\bibnamefont {Zhu}}, \bibinfo
  {author} {\bibfnamefont {Z.}~\bibnamefont {Bai}}, \bibinfo {author}
  {\bibfnamefont {Z.}~\bibnamefont {Wang}}, \bibinfo {author} {\bibfnamefont
  {S.}~\bibnamefont {Hu}}, \bibinfo {author} {\bibfnamefont {H.-Y.}\
  \bibnamefont {Xie}}, \bibinfo {author} {\bibfnamefont {X.}~\bibnamefont
  {Hu}}, \bibinfo {author} {\bibfnamefont {J.}~\bibnamefont {Cui}}, \bibinfo
  {author} {\bibfnamefont {M.}~\bibnamefont {Huang}}, \bibinfo {author}
  {\bibfnamefont {J.}~\bibnamefont {Chen}}, \bibinfo {author} {\bibfnamefont
  {Y.}~\bibnamefont {Ding}}, \bibinfo {author} {\bibfnamefont {L.}~\bibnamefont
  {Zhao}}, \bibinfo {author} {\bibfnamefont {X.}~\bibnamefont {Li}}, \bibinfo
  {author} {\bibfnamefont {Q.}~\bibnamefont {Zhang}}, \bibinfo {author}
  {\bibfnamefont {L.}~\bibnamefont {Gu}}, \bibinfo {author} {\bibfnamefont
  {X.~J.}\ \bibnamefont {Zhou}}, \bibinfo {author} {\bibfnamefont
  {J.}~\bibnamefont {Zhu}}, \bibinfo {author} {\bibfnamefont {D.}~\bibnamefont
  {Zhang}},\ and\ \bibinfo {author} {\bibfnamefont {Q.-K.}\ \bibnamefont
  {Xue}},\ }\bibfield  {title} {\bibinfo {title} {Prominent josephson tunneling
  between twisted single copper oxide planes of bi2sr2-xlaxcuo6+y},\ }\href
  {https://doi.org/10.1038/s41467-023-40525-1} {\bibfield  {journal} {\bibinfo
  {journal} {Nature Communications}\ }\textbf {\bibinfo {volume} {14}},\
  \bibinfo {pages} {5201} (\bibinfo {year} {2023})}\BibitemShut {NoStop}%
\bibitem [{\citenamefont {Song}\ \emph {et~al.}(2022)\citenamefont {Song},
  \citenamefont {Zhang},\ and\ \citenamefont {Vishwanath}}]{so.zh.22}%
  \BibitemOpen
  \bibfield  {author} {\bibinfo {author} {\bibfnamefont {X.-Y.}\ \bibnamefont
  {Song}}, \bibinfo {author} {\bibfnamefont {Y.-H.}\ \bibnamefont {Zhang}},\
  and\ \bibinfo {author} {\bibfnamefont {A.}~\bibnamefont {Vishwanath}},\
  }\bibfield  {title} {\bibinfo {title} {Doping a moir\'e mott insulator: A
  $t\ensuremath{-}j$ model study of twisted cuprates},\ }\href
  {https://doi.org/10.1103/PhysRevB.105.L201102} {\bibfield  {journal}
  {\bibinfo  {journal} {Phys. Rev. B}\ }\textbf {\bibinfo {volume} {105}},\
  \bibinfo {pages} {L201102} (\bibinfo {year} {2022})}\BibitemShut {NoStop}%
\bibitem [{\citenamefont {Lu}\ and\ \citenamefont
  {S\'en\'echal}(2022)}]{lu.se.22}%
  \BibitemOpen
  \bibfield  {author} {\bibinfo {author} {\bibfnamefont {X.}~\bibnamefont
  {Lu}}\ and\ \bibinfo {author} {\bibfnamefont {D.}~\bibnamefont
  {S\'en\'echal}},\ }\bibfield  {title} {\bibinfo {title} {Doping phase diagram
  of a hubbard model for twisted bilayer cuprates},\ }\href
  {https://doi.org/10.1103/PhysRevB.105.245127} {\bibfield  {journal} {\bibinfo
   {journal} {Phys. Rev. B}\ }\textbf {\bibinfo {volume} {105}},\ \bibinfo
  {pages} {245127} (\bibinfo {year} {2022})}\BibitemShut {NoStop}%
\bibitem [{\citenamefont {Liu}\ \emph {et~al.}(2023)\citenamefont {Liu},
  \citenamefont {Zhang}, \citenamefont {Chen},\ and\ \citenamefont
  {Yang}}]{li.zh.23}%
  \BibitemOpen
  \bibfield  {author} {\bibinfo {author} {\bibfnamefont {Y.-B.}\ \bibnamefont
  {Liu}}, \bibinfo {author} {\bibfnamefont {Y.}~\bibnamefont {Zhang}}, \bibinfo
  {author} {\bibfnamefont {W.-Q.}\ \bibnamefont {Chen}},\ and\ \bibinfo
  {author} {\bibfnamefont {F.}~\bibnamefont {Yang}},\ }\bibfield  {title}
  {\bibinfo {title} {High-angular-momentum topological superconductivities in
  twisted bilayer quasicrystal systems},\ }\href
  {https://doi.org/10.1103/PhysRevB.107.014501} {\bibfield  {journal} {\bibinfo
   {journal} {Phys. Rev. B}\ }\textbf {\bibinfo {volume} {107}},\ \bibinfo
  {pages} {014501} (\bibinfo {year} {2023})}\BibitemShut {NoStop}%
\bibitem [{\citenamefont {Mercado}\ \emph {et~al.}(2022)\citenamefont
  {Mercado}, \citenamefont {Sahoo},\ and\ \citenamefont {Franz}}]{me.sa.22}%
  \BibitemOpen
  \bibfield  {author} {\bibinfo {author} {\bibfnamefont {A.}~\bibnamefont
  {Mercado}}, \bibinfo {author} {\bibfnamefont {S.}~\bibnamefont {Sahoo}},\
  and\ \bibinfo {author} {\bibfnamefont {M.}~\bibnamefont {Franz}},\ }\bibfield
   {title} {\bibinfo {title} {High-temperature majorana zero modes},\ }\href
  {https://doi.org/10.1103/PhysRevLett.128.137002} {\bibfield  {journal}
  {\bibinfo  {journal} {Phys. Rev. Lett.}\ }\textbf {\bibinfo {volume} {128}},\
  \bibinfo {pages} {137002} (\bibinfo {year} {2022})}\BibitemShut {NoStop}%
\bibitem [{\citenamefont {Margalit}\ \emph {et~al.}(2022)\citenamefont
  {Margalit}, \citenamefont {Yan}, \citenamefont {Franz},\ and\ \citenamefont
  {Oreg}}]{ma.ya.22}%
  \BibitemOpen
  \bibfield  {author} {\bibinfo {author} {\bibfnamefont {G.}~\bibnamefont
  {Margalit}}, \bibinfo {author} {\bibfnamefont {B.}~\bibnamefont {Yan}},
  \bibinfo {author} {\bibfnamefont {M.}~\bibnamefont {Franz}},\ and\ \bibinfo
  {author} {\bibfnamefont {Y.}~\bibnamefont {Oreg}},\ }\bibfield  {title}
  {\bibinfo {title} {Chiral majorana modes via proximity to a twisted cuprate
  bilayer},\ }\href {https://doi.org/10.1103/PhysRevB.106.205424} {\bibfield
  {journal} {\bibinfo  {journal} {Phys. Rev. B}\ }\textbf {\bibinfo {volume}
  {106}},\ \bibinfo {pages} {205424} (\bibinfo {year} {2022})}\BibitemShut
  {NoStop}%
\bibitem [{\citenamefont {Gotlieb}\ \emph {et~al.}(2018)\citenamefont
  {Gotlieb}, \citenamefont {Lin}, \citenamefont {Serbyn}, \citenamefont
  {Zhang}, \citenamefont {Smallwood}, \citenamefont {Jozwiak}, \citenamefont
  {Eisaki}, \citenamefont {Hussain}, \citenamefont {Vishwanath},\ and\
  \citenamefont {Lanzara}}]{go.li.18}%
  \BibitemOpen
  \bibfield  {author} {\bibinfo {author} {\bibfnamefont {K.}~\bibnamefont
  {Gotlieb}}, \bibinfo {author} {\bibfnamefont {C.-Y.}\ \bibnamefont {Lin}},
  \bibinfo {author} {\bibfnamefont {M.}~\bibnamefont {Serbyn}}, \bibinfo
  {author} {\bibfnamefont {W.}~\bibnamefont {Zhang}}, \bibinfo {author}
  {\bibfnamefont {C.~L.}\ \bibnamefont {Smallwood}}, \bibinfo {author}
  {\bibfnamefont {C.}~\bibnamefont {Jozwiak}}, \bibinfo {author} {\bibfnamefont
  {H.}~\bibnamefont {Eisaki}}, \bibinfo {author} {\bibfnamefont
  {Z.}~\bibnamefont {Hussain}}, \bibinfo {author} {\bibfnamefont
  {A.}~\bibnamefont {Vishwanath}},\ and\ \bibinfo {author} {\bibfnamefont
  {A.}~\bibnamefont {Lanzara}},\ }\bibfield  {title} {\bibinfo {title}
  {Revealing hidden spin-momentum locking in a high-temperature cuprate
  superconductor},\ }\href {https://doi.org/10.1126/science.aao0980} {\bibfield
   {journal} {\bibinfo  {journal} {Science}\ }\textbf {\bibinfo {volume}
  {362}},\ \bibinfo {pages} {1271} (\bibinfo {year} {2018})},\ \Eprint
  {https://arxiv.org/abs/https://science.sciencemag.org/content/362/6420/1271.full.pdf}
  {https://science.sciencemag.org/content/362/6420/1271.full.pdf} \BibitemShut
  {NoStop}%
\bibitem [{\citenamefont {Fischer}\ \emph {et~al.}(2011)\citenamefont
  {Fischer}, \citenamefont {Loder},\ and\ \citenamefont {Sigrist}}]{fi.lo.11}%
  \BibitemOpen
  \bibfield  {author} {\bibinfo {author} {\bibfnamefont {M.~H.}\ \bibnamefont
  {Fischer}}, \bibinfo {author} {\bibfnamefont {F.}~\bibnamefont {Loder}},\
  and\ \bibinfo {author} {\bibfnamefont {M.}~\bibnamefont {Sigrist}},\
  }\bibfield  {title} {\bibinfo {title} {Superconductivity and local
  noncentrosymmetricity in crystal lattices},\ }\href
  {https://doi.org/10.1103/PhysRevB.84.184533} {\bibfield  {journal} {\bibinfo
  {journal} {Phys. Rev. B}\ }\textbf {\bibinfo {volume} {84}},\ \bibinfo
  {pages} {184533} (\bibinfo {year} {2011})}\BibitemShut {NoStop}%
\bibitem [{\citenamefont {Maruyama}\ \emph {et~al.}(2012)\citenamefont
  {Maruyama}, \citenamefont {Sigrist},\ and\ \citenamefont
  {Yanase}}]{ma.si.12}%
  \BibitemOpen
  \bibfield  {author} {\bibinfo {author} {\bibfnamefont {D.}~\bibnamefont
  {Maruyama}}, \bibinfo {author} {\bibfnamefont {M.}~\bibnamefont {Sigrist}},\
  and\ \bibinfo {author} {\bibfnamefont {Y.}~\bibnamefont {Yanase}},\
  }\bibfield  {title} {\bibinfo {title} {Locally non-centrosymmetric
  superconductivity in multilayer systems},\ }\href
  {https://doi.org/10.1143/JPSJ.81.034702} {\bibfield  {journal} {\bibinfo
  {journal} {Journal of the Physical Society of Japan}\ }\textbf {\bibinfo
  {volume} {81}},\ \bibinfo {pages} {034702} (\bibinfo {year} {2012})},\
  \Eprint {https://arxiv.org/abs/https://doi.org/10.1143/JPSJ.81.034702}
  {https://doi.org/10.1143/JPSJ.81.034702} \BibitemShut {NoStop}%
\bibitem [{\citenamefont {Zhang}\ \emph {et~al.}(2014)\citenamefont {Zhang},
  \citenamefont {Liu}, \citenamefont {Luo}, \citenamefont {Freeman},\ and\
  \citenamefont {Zunger}}]{zh.li.14}%
  \BibitemOpen
  \bibfield  {author} {\bibinfo {author} {\bibfnamefont {X.}~\bibnamefont
  {Zhang}}, \bibinfo {author} {\bibfnamefont {Q.}~\bibnamefont {Liu}}, \bibinfo
  {author} {\bibfnamefont {J.-W.}\ \bibnamefont {Luo}}, \bibinfo {author}
  {\bibfnamefont {A.~J.}\ \bibnamefont {Freeman}},\ and\ \bibinfo {author}
  {\bibfnamefont {A.}~\bibnamefont {Zunger}},\ }\bibfield  {title} {\bibinfo
  {title} {Hidden spin polarization in inversion-symmetric bulk crystals},\
  }\href {https://doi.org/10.1038/nphys2933} {\bibfield  {journal} {\bibinfo
  {journal} {Nature Physics}\ }\textbf {\bibinfo {volume} {10}},\ \bibinfo
  {pages} {387} (\bibinfo {year} {2014})}\BibitemShut {NoStop}%
\bibitem [{\citenamefont {Lu}\ and\ \citenamefont
  {S\'en\'echal}(2021)}]{lu.se.21}%
  \BibitemOpen
  \bibfield  {author} {\bibinfo {author} {\bibfnamefont {X.}~\bibnamefont
  {Lu}}\ and\ \bibinfo {author} {\bibfnamefont {D.}~\bibnamefont
  {S\'en\'echal}},\ }\bibfield  {title} {\bibinfo {title} {Spin texture in a
  bilayer high-temperature cuprate superconductor},\ }\href
  {https://doi.org/10.1103/PhysRevB.104.024502} {\bibfield  {journal} {\bibinfo
   {journal} {Phys. Rev. B}\ }\textbf {\bibinfo {volume} {104}},\ \bibinfo
  {pages} {024502} (\bibinfo {year} {2021})}\BibitemShut {NoStop}%
\bibitem [{\citenamefont {Fischer}\ \emph {et~al.}(2023)\citenamefont
  {Fischer}, \citenamefont {Sigrist}, \citenamefont {Agterberg},\ and\
  \citenamefont {Yanase}}]{fi.si.22}%
  \BibitemOpen
  \bibfield  {author} {\bibinfo {author} {\bibfnamefont {M.~H.}\ \bibnamefont
  {Fischer}}, \bibinfo {author} {\bibfnamefont {M.}~\bibnamefont {Sigrist}},
  \bibinfo {author} {\bibfnamefont {D.~F.}\ \bibnamefont {Agterberg}},\ and\
  \bibinfo {author} {\bibfnamefont {Y.}~\bibnamefont {Yanase}},\ }\bibfield
  {title} {\bibinfo {title} {Superconductivity and local inversion-symmetry
  breaking},\ }\href {https://doi.org/10.1146/annurev-conmatphys-040521-042511}
  {\bibfield  {journal} {\bibinfo  {journal} {Annual Review of Condensed Matter
  Physics}\ }\textbf {\bibinfo {volume} {14}},\ \bibinfo {pages} {null}
  (\bibinfo {year} {2023})},\ \Eprint
  {https://arxiv.org/abs/https://doi.org/10.1146/annurev-conmatphys-040521-042511}
  {https://doi.org/10.1146/annurev-conmatphys-040521-042511} \BibitemShut
  {NoStop}%
\bibitem [{\citenamefont {Sato}\ \emph {et~al.}(2010)\citenamefont {Sato},
  \citenamefont {Takahashi},\ and\ \citenamefont {Fujimoto}}]{sa.ta.10}%
  \BibitemOpen
  \bibfield  {author} {\bibinfo {author} {\bibfnamefont {M.}~\bibnamefont
  {Sato}}, \bibinfo {author} {\bibfnamefont {Y.}~\bibnamefont {Takahashi}},\
  and\ \bibinfo {author} {\bibfnamefont {S.}~\bibnamefont {Fujimoto}},\
  }\bibfield  {title} {\bibinfo {title} {Non-abelian topological orders and
  majorana fermions in spin-singlet superconductors},\ }\href@noop {}
  {\bibfield  {journal} {\bibinfo  {journal} {Physical Review B}\ }\textbf
  {\bibinfo {volume} {82}},\ \bibinfo {pages} {134521} (\bibinfo {year}
  {2010})}\BibitemShut {NoStop}%
\bibitem [{\citenamefont {Sato}\ and\ \citenamefont
  {Fujimoto}(2010)}]{sa.fu.10}%
  \BibitemOpen
  \bibfield  {author} {\bibinfo {author} {\bibfnamefont {M.}~\bibnamefont
  {Sato}}\ and\ \bibinfo {author} {\bibfnamefont {S.}~\bibnamefont
  {Fujimoto}},\ }\bibfield  {title} {\bibinfo {title} {Existence of majorana
  fermions and topological order in nodal superconductors with spin-orbit
  interactions in external magnetic fields},\ }\href
  {https://doi.org/10.1103/PhysRevLett.105.217001} {\bibfield  {journal}
  {\bibinfo  {journal} {Phys. Rev. Lett.}\ }\textbf {\bibinfo {volume} {105}},\
  \bibinfo {pages} {217001} (\bibinfo {year} {2010})}\BibitemShut {NoStop}%
\bibitem [{\citenamefont {Volkov}\ \emph {et~al.}(2021)\citenamefont {Volkov},
  \citenamefont {Zhao}, \citenamefont {Poccia}, \citenamefont {Cui},
  \citenamefont {Kim},\ and\ \citenamefont {Pixley}}]{vo.zh.21}%
  \BibitemOpen
  \bibfield  {author} {\bibinfo {author} {\bibfnamefont {P.~A.}\ \bibnamefont
  {Volkov}}, \bibinfo {author} {\bibfnamefont {S.~Y.~F.}\ \bibnamefont {Zhao}},
  \bibinfo {author} {\bibfnamefont {N.}~\bibnamefont {Poccia}}, \bibinfo
  {author} {\bibfnamefont {X.}~\bibnamefont {Cui}}, \bibinfo {author}
  {\bibfnamefont {P.}~\bibnamefont {Kim}},\ and\ \bibinfo {author}
  {\bibfnamefont {J.~H.}\ \bibnamefont {Pixley}},\ }\href@noop {} {\bibinfo
  {title} {Josephson effects in twisted nodal superconductors}} (\bibinfo
  {year} {2021}),\ \Eprint {https://arxiv.org/abs/2108.13456} {arXiv:2108.13456
  [cond-mat.supr-con]} \BibitemShut {NoStop}%
\bibitem [{\citenamefont {Yoshida}\ and\ \citenamefont
  {Yanase}(2016)}]{yo.ya.16}%
  \BibitemOpen
  \bibfield  {author} {\bibinfo {author} {\bibfnamefont {T.}~\bibnamefont
  {Yoshida}}\ and\ \bibinfo {author} {\bibfnamefont {Y.}~\bibnamefont
  {Yanase}},\ }\bibfield  {title} {\bibinfo {title} {Topological $d+p$-wave
  superconductivity in rashba systems},\ }\href
  {https://doi.org/10.1103/PhysRevB.93.054504} {\bibfield  {journal} {\bibinfo
  {journal} {Phys. Rev. B}\ }\textbf {\bibinfo {volume} {93}},\ \bibinfo
  {pages} {054504} (\bibinfo {year} {2016})}\BibitemShut {NoStop}%
\bibitem [{\citenamefont {Volkov}\ \emph
  {et~al.}(2023{\natexlab{a}})\citenamefont {Volkov}, \citenamefont {Wilson},
  \citenamefont {Lucht},\ and\ \citenamefont {Pixley}}]{vo.wi.23a}%
  \BibitemOpen
  \bibfield  {author} {\bibinfo {author} {\bibfnamefont {P.~A.}\ \bibnamefont
  {Volkov}}, \bibinfo {author} {\bibfnamefont {J.~H.}\ \bibnamefont {Wilson}},
  \bibinfo {author} {\bibfnamefont {K.~P.}\ \bibnamefont {Lucht}},\ and\
  \bibinfo {author} {\bibfnamefont {J.~H.}\ \bibnamefont {Pixley}},\ }\bibfield
   {title} {\bibinfo {title} {Current- and field-induced topology in twisted
  nodal superconductors},\ }\href
  {https://doi.org/10.1103/PhysRevLett.130.186001} {\bibfield  {journal}
  {\bibinfo  {journal} {Phys. Rev. Lett.}\ }\textbf {\bibinfo {volume} {130}},\
  \bibinfo {pages} {186001} (\bibinfo {year} {2023}{\natexlab{a}})}\BibitemShut
  {NoStop}%
\bibitem [{\citenamefont {Volkov}\ \emph
  {et~al.}(2023{\natexlab{b}})\citenamefont {Volkov}, \citenamefont {Wilson},
  \citenamefont {Lucht},\ and\ \citenamefont {Pixley}}]{vo.wi.23b}%
  \BibitemOpen
  \bibfield  {author} {\bibinfo {author} {\bibfnamefont {P.~A.}\ \bibnamefont
  {Volkov}}, \bibinfo {author} {\bibfnamefont {J.~H.}\ \bibnamefont {Wilson}},
  \bibinfo {author} {\bibfnamefont {K.~P.}\ \bibnamefont {Lucht}},\ and\
  \bibinfo {author} {\bibfnamefont {J.~H.}\ \bibnamefont {Pixley}},\ }\bibfield
   {title} {\bibinfo {title} {Magic angles and correlations in twisted nodal
  superconductors},\ }\href {https://doi.org/10.1103/PhysRevB.107.174506}
  {\bibfield  {journal} {\bibinfo  {journal} {Phys. Rev. B}\ }\textbf {\bibinfo
  {volume} {107}},\ \bibinfo {pages} {174506} (\bibinfo {year}
  {2023}{\natexlab{b}})}\BibitemShut {NoStop}%
\bibitem [{\citenamefont {Moon}\ and\ \citenamefont
  {Koshino}(2013)}]{mo.ko.13}%
  \BibitemOpen
  \bibfield  {author} {\bibinfo {author} {\bibfnamefont {P.}~\bibnamefont
  {Moon}}\ and\ \bibinfo {author} {\bibfnamefont {M.}~\bibnamefont {Koshino}},\
  }\bibfield  {title} {\bibinfo {title} {Optical absorption in twisted bilayer
  graphene},\ }\href {https://doi.org/10.1103/PhysRevB.87.205404} {\bibfield
  {journal} {\bibinfo  {journal} {Phys. Rev. B}\ }\textbf {\bibinfo {volume}
  {87}},\ \bibinfo {pages} {205404} (\bibinfo {year} {2013})}\BibitemShut
  {NoStop}%
\bibitem [{\citenamefont {Koshino}(2015)}]{kosh.15}%
  \BibitemOpen
  \bibfield  {author} {\bibinfo {author} {\bibfnamefont {M.}~\bibnamefont
  {Koshino}},\ }\bibfield  {title} {\bibinfo {title} {Interlayer interaction in
  general incommensurate atomic layers},\ }\href
  {https://doi.org/10.1088/1367-2630/17/1/015014} {\bibfield  {journal}
  {\bibinfo  {journal} {New Journal of Physics}\ }\textbf {\bibinfo {volume}
  {17}},\ \bibinfo {pages} {015014} (\bibinfo {year} {2015})}\BibitemShut
  {NoStop}%
\bibitem [{\citenamefont {Tummuru}\ \emph
  {et~al.}(2022{\natexlab{a}})\citenamefont {Tummuru}, \citenamefont {Plugge},\
  and\ \citenamefont {Franz}}]{tu.pl.22}%
  \BibitemOpen
  \bibfield  {author} {\bibinfo {author} {\bibfnamefont {T.}~\bibnamefont
  {Tummuru}}, \bibinfo {author} {\bibfnamefont {S.}~\bibnamefont {Plugge}},\
  and\ \bibinfo {author} {\bibfnamefont {M.}~\bibnamefont {Franz}},\ }\bibfield
   {title} {\bibinfo {title} {Josephson effects in twisted cuprate bilayers},\
  }\href {https://doi.org/10.1103/PhysRevB.105.064501} {\bibfield  {journal}
  {\bibinfo  {journal} {Phys. Rev. B}\ }\textbf {\bibinfo {volume} {105}},\
  \bibinfo {pages} {064501} (\bibinfo {year} {2022}{\natexlab{a}})}\BibitemShut
  {NoStop}%
\bibitem [{\citenamefont {Tummuru}\ \emph
  {et~al.}(2022{\natexlab{b}})\citenamefont {Tummuru}, \citenamefont
  {Lantagne-Hurtubise},\ and\ \citenamefont {Franz}}]{tu.la.22}%
  \BibitemOpen
  \bibfield  {author} {\bibinfo {author} {\bibfnamefont {T.}~\bibnamefont
  {Tummuru}}, \bibinfo {author} {\bibfnamefont {E.}~\bibnamefont
  {Lantagne-Hurtubise}},\ and\ \bibinfo {author} {\bibfnamefont
  {M.}~\bibnamefont {Franz}},\ }\bibfield  {title} {\bibinfo {title} {Twisted
  multilayer nodal superconductors},\ }\href
  {https://doi.org/10.1103/PhysRevB.106.014520} {\bibfield  {journal} {\bibinfo
   {journal} {Phys. Rev. B}\ }\textbf {\bibinfo {volume} {106}},\ \bibinfo
  {pages} {014520} (\bibinfo {year} {2022}{\natexlab{b}})}\BibitemShut
  {NoStop}%
\bibitem [{\citenamefont {Iwano}\ and\ \citenamefont
  {Yamaji}(2022)}]{iw.ya.22}%
  \BibitemOpen
  \bibfield  {author} {\bibinfo {author} {\bibfnamefont {A.}~\bibnamefont
  {Iwano}}\ and\ \bibinfo {author} {\bibfnamefont {Y.}~\bibnamefont {Yamaji}},\
  }\bibfield  {title} {\bibinfo {title} {Superconductivity in bilayer t–t'
  hubbard models},\ }\href {https://doi.org/10.7566/JPSJ.91.094702} {\bibfield
  {journal} {\bibinfo  {journal} {Journal of the Physical Society of Japan}\
  }\textbf {\bibinfo {volume} {91}},\ \bibinfo {pages} {094702} (\bibinfo
  {year} {2022})},\ \Eprint
  {https://arxiv.org/abs/https://doi.org/10.7566/JPSJ.91.094702}
  {https://doi.org/10.7566/JPSJ.91.094702} \BibitemShut {NoStop}%
\bibitem [{\citenamefont {Scalettar}\ \emph {et~al.}(1994)\citenamefont
  {Scalettar}, \citenamefont {Cannon}, \citenamefont {Scalapino},\ and\
  \citenamefont {Sugar}}]{sc.ca.94}%
  \BibitemOpen
  \bibfield  {author} {\bibinfo {author} {\bibfnamefont {R.~T.}\ \bibnamefont
  {Scalettar}}, \bibinfo {author} {\bibfnamefont {J.~W.}\ \bibnamefont
  {Cannon}}, \bibinfo {author} {\bibfnamefont {D.~J.}\ \bibnamefont
  {Scalapino}},\ and\ \bibinfo {author} {\bibfnamefont {R.~L.}\ \bibnamefont
  {Sugar}},\ }\bibfield  {title} {\bibinfo {title} {Magnetic and pairing
  correlations in coupled hubbard planes},\ }\href
  {https://doi.org/10.1103/PhysRevB.50.13419} {\bibfield  {journal} {\bibinfo
  {journal} {Phys. Rev. B}\ }\textbf {\bibinfo {volume} {50}},\ \bibinfo
  {pages} {13419} (\bibinfo {year} {1994})}\BibitemShut {NoStop}%
\bibitem [{\citenamefont {Haenel}\ \emph {et~al.}(2022)\citenamefont {Haenel},
  \citenamefont {Tummuru},\ and\ \citenamefont {Franz}}]{ha.tu.22}%
  \BibitemOpen
  \bibfield  {author} {\bibinfo {author} {\bibfnamefont {R.}~\bibnamefont
  {Haenel}}, \bibinfo {author} {\bibfnamefont {T.}~\bibnamefont {Tummuru}},\
  and\ \bibinfo {author} {\bibfnamefont {M.}~\bibnamefont {Franz}},\ }\bibfield
   {title} {\bibinfo {title} {Incoherent tunneling and topological
  superconductivity in twisted cuprate bilayers},\ }\href
  {https://doi.org/10.1103/PhysRevB.106.104505} {\bibfield  {journal} {\bibinfo
   {journal} {Phys. Rev. B}\ }\textbf {\bibinfo {volume} {106}},\ \bibinfo
  {pages} {104505} (\bibinfo {year} {2022})}\BibitemShut {NoStop}%
\bibitem [{\citenamefont {Sato}\ \emph {et~al.}(2009)\citenamefont {Sato},
  \citenamefont {Takahashi},\ and\ \citenamefont {Fujimoto}}]{sa.ta.09}%
  \BibitemOpen
  \bibfield  {author} {\bibinfo {author} {\bibfnamefont {M.}~\bibnamefont
  {Sato}}, \bibinfo {author} {\bibfnamefont {Y.}~\bibnamefont {Takahashi}},\
  and\ \bibinfo {author} {\bibfnamefont {S.}~\bibnamefont {Fujimoto}},\
  }\bibfield  {title} {\bibinfo {title} {Non-abelian topological order in
  $s$-wave superfluids of ultracold fermionic atoms},\ }\href
  {https://doi.org/10.1103/PhysRevLett.103.020401} {\bibfield  {journal}
  {\bibinfo  {journal} {Phys. Rev. Lett.}\ }\textbf {\bibinfo {volume} {103}},\
  \bibinfo {pages} {020401} (\bibinfo {year} {2009})}\BibitemShut {NoStop}%
\bibitem [{\citenamefont {Sato}(2010)}]{sato.10}%
  \BibitemOpen
  \bibfield  {author} {\bibinfo {author} {\bibfnamefont {M.}~\bibnamefont
  {Sato}},\ }\bibfield  {title} {\bibinfo {title} {Topological odd-parity
  superconductors},\ }\href {https://doi.org/10.1103/PhysRevB.81.220504}
  {\bibfield  {journal} {\bibinfo  {journal} {Phys. Rev. B}\ }\textbf {\bibinfo
  {volume} {81}},\ \bibinfo {pages} {220504} (\bibinfo {year}
  {2010})}\BibitemShut {NoStop}%
\bibitem [{\citenamefont {Fukui}\ \emph {et~al.}(2005)\citenamefont {Fukui},
  \citenamefont {Hatsugai},\ and\ \citenamefont {Suzuki}}]{fu.ha.05}%
  \BibitemOpen
  \bibfield  {author} {\bibinfo {author} {\bibfnamefont {T.}~\bibnamefont
  {Fukui}}, \bibinfo {author} {\bibfnamefont {Y.}~\bibnamefont {Hatsugai}},\
  and\ \bibinfo {author} {\bibfnamefont {H.}~\bibnamefont {Suzuki}},\
  }\bibfield  {title} {\bibinfo {title} {Chern numbers in discretized brillouin
  zone: Efficient method of computing (spin) hall conductances},\ }\href
  {https://doi.org/10.1143/JPSJ.74.1674} {\bibfield  {journal} {\bibinfo
  {journal} {Journal of the Physical Society of Japan}\ }\textbf {\bibinfo
  {volume} {74}},\ \bibinfo {pages} {1674} (\bibinfo {year}
  {2005})}\BibitemShut {NoStop}%
\bibitem [{\citenamefont {Zhu}(2019)}]{zhu.19}%
  \BibitemOpen
  \bibfield  {author} {\bibinfo {author} {\bibfnamefont {X.}~\bibnamefont
  {Zhu}},\ }\bibfield  {title} {\bibinfo {title} {Second-order topological
  superconductors with mixed pairing},\ }\href
  {https://doi.org/10.1103/PhysRevLett.122.236401} {\bibfield  {journal}
  {\bibinfo  {journal} {Phys. Rev. Lett.}\ }\textbf {\bibinfo {volume} {122}},\
  \bibinfo {pages} {236401} (\bibinfo {year} {2019})}\BibitemShut {NoStop}%
\bibitem [{\citenamefont {Yan}(2019)}]{yan.19}%
  \BibitemOpen
  \bibfield  {author} {\bibinfo {author} {\bibfnamefont {Z.}~\bibnamefont
  {Yan}},\ }\bibfield  {title} {\bibinfo {title} {Higher-order topological
  odd-parity superconductors},\ }\href
  {https://doi.org/10.1103/PhysRevLett.123.177001} {\bibfield  {journal}
  {\bibinfo  {journal} {Phys. Rev. Lett.}\ }\textbf {\bibinfo {volume} {123}},\
  \bibinfo {pages} {177001} (\bibinfo {year} {2019})}\BibitemShut {NoStop}%
\bibitem [{\citenamefont {Kheirkhah}\ \emph {et~al.}(2020)\citenamefont
  {Kheirkhah}, \citenamefont {Yan}, \citenamefont {Nagai},\ and\ \citenamefont
  {Marsiglio}}]{kh.ya.20}%
  \BibitemOpen
  \bibfield  {author} {\bibinfo {author} {\bibfnamefont {M.}~\bibnamefont
  {Kheirkhah}}, \bibinfo {author} {\bibfnamefont {Z.}~\bibnamefont {Yan}},
  \bibinfo {author} {\bibfnamefont {Y.}~\bibnamefont {Nagai}},\ and\ \bibinfo
  {author} {\bibfnamefont {F.}~\bibnamefont {Marsiglio}},\ }\bibfield  {title}
  {\bibinfo {title} {First- and second-order topological superconductivity and
  temperature-driven topological phase transitions in the extended hubbard
  model with spin-orbit coupling},\ }\href
  {https://doi.org/10.1103/PhysRevLett.125.017001} {\bibfield  {journal}
  {\bibinfo  {journal} {Phys. Rev. Lett.}\ }\textbf {\bibinfo {volume} {125}},\
  \bibinfo {pages} {017001} (\bibinfo {year} {2020})}\BibitemShut {NoStop}%
\bibitem [{\citenamefont {Kang}\ \emph {et~al.}(2019)\citenamefont {Kang},
  \citenamefont {Shiozaki},\ and\ \citenamefont {Cho}}]{ka.sh.19}%
  \BibitemOpen
  \bibfield  {author} {\bibinfo {author} {\bibfnamefont {B.}~\bibnamefont
  {Kang}}, \bibinfo {author} {\bibfnamefont {K.}~\bibnamefont {Shiozaki}},\
  and\ \bibinfo {author} {\bibfnamefont {G.~Y.}\ \bibnamefont {Cho}},\
  }\bibfield  {title} {\bibinfo {title} {Many-body order parameters for
  multipoles in solids},\ }\href {https://doi.org/10.1103/PhysRevB.100.245134}
  {\bibfield  {journal} {\bibinfo  {journal} {Phys. Rev. B}\ }\textbf {\bibinfo
  {volume} {100}},\ \bibinfo {pages} {245134} (\bibinfo {year}
  {2019})}\BibitemShut {NoStop}%
\bibitem [{\citenamefont {Wheeler}\ \emph {et~al.}(2019)\citenamefont
  {Wheeler}, \citenamefont {Wagner},\ and\ \citenamefont {Hughes}}]{wh.wa.19}%
  \BibitemOpen
  \bibfield  {author} {\bibinfo {author} {\bibfnamefont {W.~A.}\ \bibnamefont
  {Wheeler}}, \bibinfo {author} {\bibfnamefont {L.~K.}\ \bibnamefont
  {Wagner}},\ and\ \bibinfo {author} {\bibfnamefont {T.~L.}\ \bibnamefont
  {Hughes}},\ }\bibfield  {title} {\bibinfo {title} {Many-body electric
  multipole operators in extended systems},\ }\href
  {https://doi.org/10.1103/PhysRevB.100.245135} {\bibfield  {journal} {\bibinfo
   {journal} {Phys. Rev. B}\ }\textbf {\bibinfo {volume} {100}},\ \bibinfo
  {pages} {245135} (\bibinfo {year} {2019})}\BibitemShut {NoStop}%
\bibitem [{\citenamefont {Li}\ \emph {et~al.}(2020)\citenamefont {Li},
  \citenamefont {Fu}, \citenamefont {Hu}, \citenamefont {Li},\ and\
  \citenamefont {Shen}}]{li.fu.20}%
  \BibitemOpen
  \bibfield  {author} {\bibinfo {author} {\bibfnamefont {C.-A.}\ \bibnamefont
  {Li}}, \bibinfo {author} {\bibfnamefont {B.}~\bibnamefont {Fu}}, \bibinfo
  {author} {\bibfnamefont {Z.-A.}\ \bibnamefont {Hu}}, \bibinfo {author}
  {\bibfnamefont {J.}~\bibnamefont {Li}},\ and\ \bibinfo {author}
  {\bibfnamefont {S.-Q.}\ \bibnamefont {Shen}},\ }\bibfield  {title} {\bibinfo
  {title} {Topological phase transitions in disordered electric quadrupole
  insulators},\ }\href {https://doi.org/10.1103/PhysRevLett.125.166801}
  {\bibfield  {journal} {\bibinfo  {journal} {Phys. Rev. Lett.}\ }\textbf
  {\bibinfo {volume} {125}},\ \bibinfo {pages} {166801} (\bibinfo {year}
  {2020})}\BibitemShut {NoStop}%
\bibitem [{\citenamefont {Yang}\ \emph {et~al.}(2021)\citenamefont {Yang},
  \citenamefont {Li}, \citenamefont {Duan},\ and\ \citenamefont
  {Xu}}]{ya.li.21}%
  \BibitemOpen
  \bibfield  {author} {\bibinfo {author} {\bibfnamefont {Y.-B.}\ \bibnamefont
  {Yang}}, \bibinfo {author} {\bibfnamefont {K.}~\bibnamefont {Li}}, \bibinfo
  {author} {\bibfnamefont {L.-M.}\ \bibnamefont {Duan}},\ and\ \bibinfo
  {author} {\bibfnamefont {Y.}~\bibnamefont {Xu}},\ }\bibfield  {title}
  {\bibinfo {title} {Higher-order topological anderson insulators},\ }\href
  {https://doi.org/10.1103/PhysRevB.103.085408} {\bibfield  {journal} {\bibinfo
   {journal} {Phys. Rev. B}\ }\textbf {\bibinfo {volume} {103}},\ \bibinfo
  {pages} {085408} (\bibinfo {year} {2021})}\BibitemShut {NoStop}%
\bibitem [{\citenamefont {Lin}\ \emph {et~al.}(2023)\citenamefont {Lin},
  \citenamefont {Huang},\ and\ \citenamefont {Lu}}]{li.lu.23}%
  \BibitemOpen
  \bibfield  {author} {\bibinfo {author} {\bibfnamefont {C.}~\bibnamefont
  {Lin}}, \bibinfo {author} {\bibfnamefont {C.}~\bibnamefont {Huang}},\ and\
  \bibinfo {author} {\bibfnamefont {X.}~\bibnamefont {Lu}},\ }\bibfield
  {title} {\bibinfo {title} {Customizing topological phases in the twisted
  bilayer superconductors with even-parity pairings},\ }\href
  {https://doi.org/10.1088/1674-1056/acd3e3} {\bibfield  {journal} {\bibinfo
  {journal} {Chinese Physics B}\ }\textbf {\bibinfo {volume} {32}},\ \bibinfo
  {pages} {087401} (\bibinfo {year} {2023})}\BibitemShut {NoStop}%
\bibitem [{\citenamefont {Senthil}\ \emph {et~al.}(1999)\citenamefont
  {Senthil}, \citenamefont {Marston},\ and\ \citenamefont {Fisher}}]{se.ma.99}%
  \BibitemOpen
  \bibfield  {author} {\bibinfo {author} {\bibfnamefont {T.}~\bibnamefont
  {Senthil}}, \bibinfo {author} {\bibfnamefont {J.~B.}\ \bibnamefont
  {Marston}},\ and\ \bibinfo {author} {\bibfnamefont {M.~P.~A.}\ \bibnamefont
  {Fisher}},\ }\bibfield  {title} {\bibinfo {title} {Spin quantum hall effect
  in unconventional superconductors},\ }\href
  {https://doi.org/10.1103/PhysRevB.60.4245} {\bibfield  {journal} {\bibinfo
  {journal} {Phys. Rev. B}\ }\textbf {\bibinfo {volume} {60}},\ \bibinfo
  {pages} {4245} (\bibinfo {year} {1999})}\BibitemShut {NoStop}%
\bibitem [{\citenamefont {Law}\ \emph {et~al.}(2009)\citenamefont {Law},
  \citenamefont {Lee},\ and\ \citenamefont {Ng}}]{la.le.09}%
  \BibitemOpen
  \bibfield  {author} {\bibinfo {author} {\bibfnamefont {K.~T.}\ \bibnamefont
  {Law}}, \bibinfo {author} {\bibfnamefont {P.~A.}\ \bibnamefont {Lee}},\ and\
  \bibinfo {author} {\bibfnamefont {T.~K.}\ \bibnamefont {Ng}},\ }\bibfield
  {title} {\bibinfo {title} {Majorana fermion induced resonant andreev
  reflection},\ }\href {https://doi.org/10.1103/PhysRevLett.103.237001}
  {\bibfield  {journal} {\bibinfo  {journal} {Phys. Rev. Lett.}\ }\textbf
  {\bibinfo {volume} {103}},\ \bibinfo {pages} {237001} (\bibinfo {year}
  {2009})}\BibitemShut {NoStop}%
\bibitem [{\citenamefont {Liu}\ \emph {et~al.}(2012)\citenamefont {Liu},
  \citenamefont {Zhang}, \citenamefont {Mou}, \citenamefont {He}, \citenamefont
  {Ou}, \citenamefont {Wang}, \citenamefont {Li}, \citenamefont {Wang},
  \citenamefont {Zhao}, \citenamefont {He}, \citenamefont {Peng}, \citenamefont
  {Liu}, \citenamefont {Chen}, \citenamefont {Yu}, \citenamefont {Liu},
  \citenamefont {Dong}, \citenamefont {Zhang}, \citenamefont {Chen},
  \citenamefont {Xu}, \citenamefont {Hu}, \citenamefont {Chen}, \citenamefont
  {Ma}, \citenamefont {Xue},\ and\ \citenamefont {Zhou}}]{li.zh.12}%
  \BibitemOpen
  \bibfield  {author} {\bibinfo {author} {\bibfnamefont {D.}~\bibnamefont
  {Liu}}, \bibinfo {author} {\bibfnamefont {W.}~\bibnamefont {Zhang}}, \bibinfo
  {author} {\bibfnamefont {D.}~\bibnamefont {Mou}}, \bibinfo {author}
  {\bibfnamefont {J.}~\bibnamefont {He}}, \bibinfo {author} {\bibfnamefont
  {Y.-B.}\ \bibnamefont {Ou}}, \bibinfo {author} {\bibfnamefont {Q.-Y.}\
  \bibnamefont {Wang}}, \bibinfo {author} {\bibfnamefont {Z.}~\bibnamefont
  {Li}}, \bibinfo {author} {\bibfnamefont {L.}~\bibnamefont {Wang}}, \bibinfo
  {author} {\bibfnamefont {L.}~\bibnamefont {Zhao}}, \bibinfo {author}
  {\bibfnamefont {S.}~\bibnamefont {He}}, \bibinfo {author} {\bibfnamefont
  {Y.}~\bibnamefont {Peng}}, \bibinfo {author} {\bibfnamefont {X.}~\bibnamefont
  {Liu}}, \bibinfo {author} {\bibfnamefont {C.}~\bibnamefont {Chen}}, \bibinfo
  {author} {\bibfnamefont {L.}~\bibnamefont {Yu}}, \bibinfo {author}
  {\bibfnamefont {G.}~\bibnamefont {Liu}}, \bibinfo {author} {\bibfnamefont
  {X.}~\bibnamefont {Dong}}, \bibinfo {author} {\bibfnamefont {J.}~\bibnamefont
  {Zhang}}, \bibinfo {author} {\bibfnamefont {C.}~\bibnamefont {Chen}},
  \bibinfo {author} {\bibfnamefont {Z.}~\bibnamefont {Xu}}, \bibinfo {author}
  {\bibfnamefont {J.}~\bibnamefont {Hu}}, \bibinfo {author} {\bibfnamefont
  {X.}~\bibnamefont {Chen}}, \bibinfo {author} {\bibfnamefont {X.}~\bibnamefont
  {Ma}}, \bibinfo {author} {\bibfnamefont {Q.}~\bibnamefont {Xue}},\ and\
  \bibinfo {author} {\bibfnamefont {X.~J.}\ \bibnamefont {Zhou}},\ }\bibfield
  {title} {\bibinfo {title} {Electronic origin of high-temperature
  superconductivity in single-layer fese superconductor},\ }\href
  {https://doi.org/10.1038/ncomms1946} {\bibfield  {journal} {\bibinfo
  {journal} {Nature Communications}\ }\textbf {\bibinfo {volume} {3}},\
  \bibinfo {pages} {931} (\bibinfo {year} {2012})}\BibitemShut {NoStop}%
\bibitem [{\citenamefont {Wu}\ \emph {et~al.}(2018)\citenamefont {Wu},
  \citenamefont {Fatemi}, \citenamefont {Gibson}, \citenamefont {Watanabe},
  \citenamefont {Taniguchi}, \citenamefont {Cava},\ and\ \citenamefont
  {Jarillo-Herrero}}]{wu.fa.18}%
  \BibitemOpen
  \bibfield  {author} {\bibinfo {author} {\bibfnamefont {S.}~\bibnamefont
  {Wu}}, \bibinfo {author} {\bibfnamefont {V.}~\bibnamefont {Fatemi}}, \bibinfo
  {author} {\bibfnamefont {Q.~D.}\ \bibnamefont {Gibson}}, \bibinfo {author}
  {\bibfnamefont {K.}~\bibnamefont {Watanabe}}, \bibinfo {author}
  {\bibfnamefont {T.}~\bibnamefont {Taniguchi}}, \bibinfo {author}
  {\bibfnamefont {R.~J.}\ \bibnamefont {Cava}},\ and\ \bibinfo {author}
  {\bibfnamefont {P.}~\bibnamefont {Jarillo-Herrero}},\ }\bibfield  {title}
  {\bibinfo {title} {Observation of the quantum spin hall effect up to 100
  kelvin in a monolayer crystal},\ }\href
  {https://doi.org/10.1126/science.aan6003} {\bibfield  {journal} {\bibinfo
  {journal} {Science}\ }\textbf {\bibinfo {volume} {359}},\ \bibinfo {pages}
  {76} (\bibinfo {year} {2018})},\ \Eprint
  {https://arxiv.org/abs/https://www.science.org/doi/pdf/10.1126/science.aan6003}
  {https://www.science.org/doi/pdf/10.1126/science.aan6003} \BibitemShut
  {NoStop}%
\bibitem [{\citenamefont {Wu}\ \emph {et~al.}(2017)\citenamefont {Wu},
  \citenamefont {Yuan}, \citenamefont {Meng}, \citenamefont {Chen},
  \citenamefont {Sun}, \citenamefont {Chen}, \citenamefont {Dang},
  \citenamefont {Tan}, \citenamefont {Liu}, \citenamefont {Yin}, \citenamefont
  {Zhou}, \citenamefont {Huang}, \citenamefont {Xu}, \citenamefont {Cui},
  \citenamefont {Hwang}, \citenamefont {Liu}, \citenamefont {Chen},
  \citenamefont {Yan},\ and\ \citenamefont {Peng}}]{wu.yu.17}%
  \BibitemOpen
  \bibfield  {author} {\bibinfo {author} {\bibfnamefont {J.}~\bibnamefont
  {Wu}}, \bibinfo {author} {\bibfnamefont {H.}~\bibnamefont {Yuan}}, \bibinfo
  {author} {\bibfnamefont {M.}~\bibnamefont {Meng}}, \bibinfo {author}
  {\bibfnamefont {C.}~\bibnamefont {Chen}}, \bibinfo {author} {\bibfnamefont
  {Y.}~\bibnamefont {Sun}}, \bibinfo {author} {\bibfnamefont {Z.}~\bibnamefont
  {Chen}}, \bibinfo {author} {\bibfnamefont {W.}~\bibnamefont {Dang}}, \bibinfo
  {author} {\bibfnamefont {C.}~\bibnamefont {Tan}}, \bibinfo {author}
  {\bibfnamefont {Y.}~\bibnamefont {Liu}}, \bibinfo {author} {\bibfnamefont
  {J.}~\bibnamefont {Yin}}, \bibinfo {author} {\bibfnamefont {Y.}~\bibnamefont
  {Zhou}}, \bibinfo {author} {\bibfnamefont {S.}~\bibnamefont {Huang}},
  \bibinfo {author} {\bibfnamefont {H.~Q.}\ \bibnamefont {Xu}}, \bibinfo
  {author} {\bibfnamefont {Y.}~\bibnamefont {Cui}}, \bibinfo {author}
  {\bibfnamefont {H.~Y.}\ \bibnamefont {Hwang}}, \bibinfo {author}
  {\bibfnamefont {Z.}~\bibnamefont {Liu}}, \bibinfo {author} {\bibfnamefont
  {Y.}~\bibnamefont {Chen}}, \bibinfo {author} {\bibfnamefont {B.}~\bibnamefont
  {Yan}},\ and\ \bibinfo {author} {\bibfnamefont {H.}~\bibnamefont {Peng}},\
  }\bibfield  {title} {\bibinfo {title} {High electron mobility and quantum
  oscillations in non-encapsulated ultrathin semiconducting bi2o2se},\ }\href
  {https://doi.org/10.1038/nnano.2017.43} {\bibfield  {journal} {\bibinfo
  {journal} {Nature Nanotechnology}\ }\textbf {\bibinfo {volume} {12}},\
  \bibinfo {pages} {530} (\bibinfo {year} {2017})}\BibitemShut {NoStop}%
\bibitem [{\citenamefont {Naritsuka}\ \emph {et~al.}(2021)\citenamefont
  {Naritsuka}, \citenamefont {Terashima},\ and\ \citenamefont
  {Matsuda}}]{na.te.21}%
  \BibitemOpen
  \bibfield  {author} {\bibinfo {author} {\bibfnamefont {M.}~\bibnamefont
  {Naritsuka}}, \bibinfo {author} {\bibfnamefont {T.}~\bibnamefont
  {Terashima}},\ and\ \bibinfo {author} {\bibfnamefont {Y.}~\bibnamefont
  {Matsuda}},\ }\bibfield  {title} {\bibinfo {title} {Controlling
  unconventional superconductivity in artificially engineered f-electron kondo
  superlattices},\ }\href {https://doi.org/10.1088/1361-648X/abfdf2} {\bibfield
   {journal} {\bibinfo  {journal} {Journal of Physics: Condensed Matter}\
  }\textbf {\bibinfo {volume} {33}},\ \bibinfo {pages} {273001} (\bibinfo
  {year} {2021})}\BibitemShut {NoStop}%
\bibitem [{\citenamefont {Mizukami}\ \emph {et~al.}(2011)\citenamefont
  {Mizukami}, \citenamefont {Shishido}, \citenamefont {Shibauchi},
  \citenamefont {Shimozawa}, \citenamefont {Yasumoto}, \citenamefont
  {Watanabe}, \citenamefont {Yamashita}, \citenamefont {Ikeda}, \citenamefont
  {Terashima}, \citenamefont {Kontani},\ and\ \citenamefont
  {Matsuda}}]{mi.sh.11}%
  \BibitemOpen
  \bibfield  {author} {\bibinfo {author} {\bibfnamefont {Y.}~\bibnamefont
  {Mizukami}}, \bibinfo {author} {\bibfnamefont {H.}~\bibnamefont {Shishido}},
  \bibinfo {author} {\bibfnamefont {T.}~\bibnamefont {Shibauchi}}, \bibinfo
  {author} {\bibfnamefont {M.}~\bibnamefont {Shimozawa}}, \bibinfo {author}
  {\bibfnamefont {S.}~\bibnamefont {Yasumoto}}, \bibinfo {author}
  {\bibfnamefont {D.}~\bibnamefont {Watanabe}}, \bibinfo {author}
  {\bibfnamefont {M.}~\bibnamefont {Yamashita}}, \bibinfo {author}
  {\bibfnamefont {H.}~\bibnamefont {Ikeda}}, \bibinfo {author} {\bibfnamefont
  {T.}~\bibnamefont {Terashima}}, \bibinfo {author} {\bibfnamefont
  {H.}~\bibnamefont {Kontani}},\ and\ \bibinfo {author} {\bibfnamefont
  {Y.}~\bibnamefont {Matsuda}},\ }\bibfield  {title} {\bibinfo {title}
  {Extremely strong-coupling superconductivity in artificial two-dimensional
  kondo lattices},\ }\href {https://doi.org/10.1038/nphys2112} {\bibfield
  {journal} {\bibinfo  {journal} {Nature Physics}\ }\textbf {\bibinfo {volume}
  {7}},\ \bibinfo {pages} {849} (\bibinfo {year} {2011})}\BibitemShut {NoStop}%
\bibitem [{\citenamefont {Naritsuka}\ \emph {et~al.}(2017)\citenamefont
  {Naritsuka}, \citenamefont {Ishii}, \citenamefont {Miyake}, \citenamefont
  {Tokiwa}, \citenamefont {Toda}, \citenamefont {Shimozawa}, \citenamefont
  {Terashima}, \citenamefont {Shibauchi}, \citenamefont {Matsuda},\ and\
  \citenamefont {Kasahara}}]{na.is.17}%
  \BibitemOpen
  \bibfield  {author} {\bibinfo {author} {\bibfnamefont {M.}~\bibnamefont
  {Naritsuka}}, \bibinfo {author} {\bibfnamefont {T.}~\bibnamefont {Ishii}},
  \bibinfo {author} {\bibfnamefont {S.}~\bibnamefont {Miyake}}, \bibinfo
  {author} {\bibfnamefont {Y.}~\bibnamefont {Tokiwa}}, \bibinfo {author}
  {\bibfnamefont {R.}~\bibnamefont {Toda}}, \bibinfo {author} {\bibfnamefont
  {M.}~\bibnamefont {Shimozawa}}, \bibinfo {author} {\bibfnamefont
  {T.}~\bibnamefont {Terashima}}, \bibinfo {author} {\bibfnamefont
  {T.}~\bibnamefont {Shibauchi}}, \bibinfo {author} {\bibfnamefont
  {Y.}~\bibnamefont {Matsuda}},\ and\ \bibinfo {author} {\bibfnamefont
  {Y.}~\bibnamefont {Kasahara}},\ }\bibfield  {title} {\bibinfo {title}
  {Emergent exotic superconductivity in artificially engineered tricolor kondo
  superlattices},\ }\href {https://doi.org/10.1103/PhysRevB.96.174512}
  {\bibfield  {journal} {\bibinfo  {journal} {Phys. Rev. B}\ }\textbf {\bibinfo
  {volume} {96}},\ \bibinfo {pages} {174512} (\bibinfo {year}
  {2017})}\BibitemShut {NoStop}%
\end{thebibliography}

%

\end{document}